\documentclass[conference]{IEEEtran}
\usepackage{fancyhdr}
\IEEEoverridecommandlockouts

\usepackage{colortbl}
\usepackage{physics}
\usepackage{url}
\usepackage[hidelinks]{hyperref}
\hypersetup{colorlinks=false, citecolor=blue, linkcolor=blue, urlcolor=blue}
\usepackage{cleveref}
\usepackage{mathtools}
\usepackage{algorithm}
\usepackage[noend]{algpseudocode}
\usepackage{cite}
\usepackage{amsmath,amssymb,amsfonts}
\usepackage{graphicx}

\usepackage{textcomp}
\usepackage{xcolor}
\usepackage{booktabs}
\usepackage{makecell}
\usepackage{subcaption}
\usepackage{etoolbox}
\usepackage{placeins}
\usepackage{tikz}
\usetikzlibrary{shapes.geometric, arrows.meta, positioning, fit, backgrounds}
\newtoggle{cameraready}

\makeatletter %
\newcommand{\linebreakand}{%
  \end{@IEEEauthorhalign}
  \hfill\mbox{}\par
  \mbox{}\hfill\begin{@IEEEauthorhalign}
}
\makeatother %

\newtheorem{definition}{Definition}

\definecolor{RED}{RGB}{190, 30, 49}
\definecolor{BLUE}{RGB}{11, 78, 179}
\definecolor{GREEN}{RGB}{0, 171, 86}
\definecolor{OLIVE}{RGB}{93, 150, 72}
\definecolor{GREY}{RGB}{50, 50, 50}
\definecolor{BROWN}{RGB}{122, 90, 40}

\DeclareMathOperator*{\argmin}{arg\,min}
\DeclareMathOperator*{\argmax}{arg\,max}

\newcommand{\mc}[1]{\mathcal{#1}}

\newcommand{\be}{\begin{equation}}
\newcommand{\ee}{\end{equation}}
\newcommand{\bes}{\begin{equation*}}
\newcommand{\ees}{\end{equation*}}

\newcommand{\paramvec}{\vb{m}}

\newcommand{\obsvec}{\vb{d}^{\text{obs}}}

\newcommand{\Kmat}{\mathbf{K}}

\newcommand{\ptomat}{\mathbf{F}}

\newcommand{\Gmat}{\mathbf{G}}

\newcommand{\numdata}{N_d}

\newcommand{\bfnoise}{\mathbf{\Gamma}_{\hskip -1pt \text{noise}}}
\newcommand{\bfpriorcov}{\mathbf{\Gamma}_{\hskip -1pt \text{prior}}}
\newcommand{\bfpostcov}{\mathbf{\Gamma}_{\hskip -1pt \text{post}}}

\newcommand{\piprior}{\pi_{\text{prior}}}
\newcommand{\pipost}{\pi_{\text{post}}}
\newcommand{\pilike}{\pi_{\text{like}}}

\newcommand{\bfmmap}{\vb{m}_{\text{map}}}

\newcommand{\bfpostcovm}{\mathbf{\Gamma}_{\hskip -1pt \text{post}}}
\newcommand{\bfpriorcovm}{\mathbf{\Gamma}_{\hskip -1pt \text{prior}}}

\toggletrue{cameraready}

\def\BibTeX{{\rm B\kern-.05em{\sc i\kern-.025em b}\kern-.08em
    T\kern-.1667em\lower.7ex\hbox{E}\kern-.125emX}}
\begin{document}

\title{
Sensor Placement for Tsunami Early Warning via\\[5pt]
Large-Scale Bayesian Optimal Experimental Design
\iftoggle{cameraready}{
}
}
\iftoggle{cameraready}{
\author{\IEEEauthorblockN{Sreeram Venkat}
\IEEEauthorblockA{
\textit{The University of Texas at Austin}\\
Austin, TX, USA \\
srvenkat@utexas.edu}
\and
\IEEEauthorblockN{Stefan Henneking}
\IEEEauthorblockA{
\textit{The University of Texas at Austin}\\
Austin, TX, USA \\
stefan@oden.utexas.edu}
\and
\IEEEauthorblockN{Omar Ghattas}
\IEEEauthorblockA{
\textit{The University of Texas at Austin}\\
Austin, TX, USA \\
omar@oden.utexas.edu}
}
}{
\author{}
}
\maketitle
\thispagestyle{fancy}
\lhead{}
\rhead{}
\chead{}
\lfoot{\footnotesize{
SC26, November 15-20, 2026, Chicago, Illinois, USA
\newline 979-8-3195-4789-7/26/\$31.00 \copyright 2026 IEEE}}
\rfoot{}
\cfoot{}
\renewcommand{\headrulewidth}{0pt}
\renewcommand{\footrulewidth}{0pt}

\begin{abstract}
Real-time tsunami early warning relies on distributed sensor networks to infer seismic sources and seafloor motion. Optimizing these networks via Bayesian optimal experimental design (OED) is exceptionally challenging for systems governed by hyperbolic partial differential equations, which lack the spectral decay required by standard low-rank approximations. We present a scalable Bayesian OED framework for linear time-invariant systems. By reformulating the inverse problem in the data space, we transform OED into dense matrix subset selection. We propose a multi-GPU, Schur-complement-update-based, greedy algorithm that solves the OED problem using a pipelined approach that fully overlaps I/O with GPU computations. Our framework achieves near-perfect weak and strong scaling across hundreds of GPUs on Perlmutter and Frontier. Applied to the 2025 Gordon Bell Prize-winning digital twin for tsunami forecasting in the Cascadia Subduction Zone, we optimize a 175-sensor network, minimizing the uncertainty of a parameter field with over one billion degrees of freedom.
\end{abstract}

\begin{IEEEkeywords}
OED, 
D-optimal design, 
Bayesian inference, 
inverse problems,
time-invariant dynamical systems,
digital twins
\end{IEEEkeywords}

\section{Introduction}\label{sec:introduction}
Tsunamis resulting from megathrust earthquakes have the potential to inflict catastrophic loss of life and severe economic damage worldwide. Traditional early warning frameworks for tsunamis \cite{Kamigaichi2009} primarily utilize seismic indicators to rapidly estimate earthquake magnitude and location \cite{Hirshorn2021}, subsequently generating tsunami forecasts using assumptions of simplified fault geometries. However, these conventional approaches struggle with near-field events where destructive waves can make landfall in minutes \cite{leveque2018cascadia}. Furthermore, they frequently fail to capture complex rupture dynamics \cite{uphoff2017sumatra}, which can yield inaccurate, delayed, or false alerts. To enhance predictive accuracy, direct assimilation of real-time offshore ocean bottom pressure and Global Navigation Satellite System (GNSS) data can be used to dynamically constrain the tsunamigenic seafloor motion, 
independent of assumptions about the fault geometry \cite{Tsushima2009, Crowell2024}.

\begin{figure}[htb]
    \centering
    \includegraphics[width=\columnwidth]
    {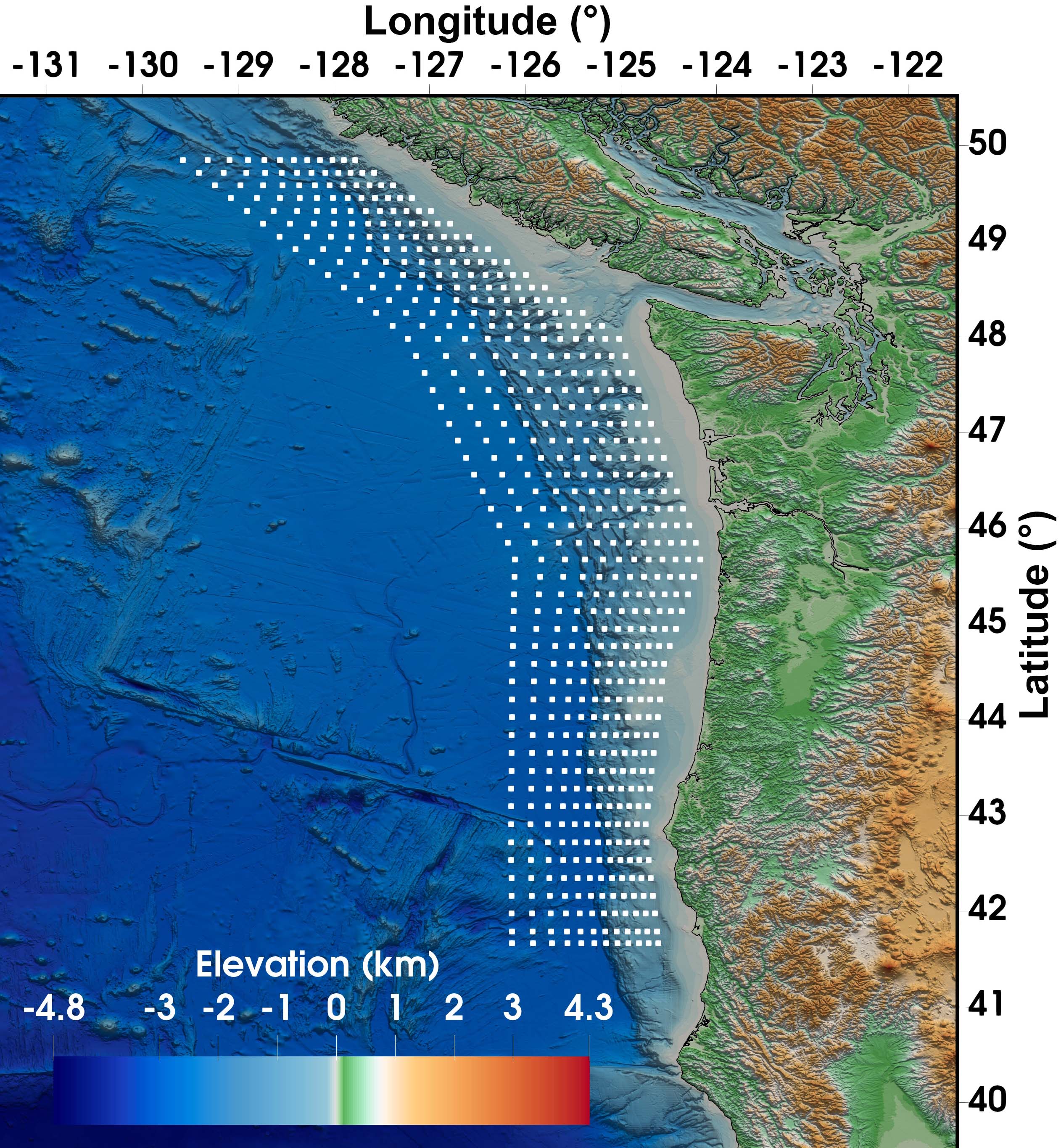}
    \caption{Topobathymetry of the Cascadia Subduction Zone with 600 candidate sensor locations.}
    \label{fig:csz-candidate-sensors}
\end{figure}

The 2025 Gordon Bell Prize-winning work \cite{henneking2025bell} showed that 
tsunami forecasting via inference 
of extreme-scale spatiotemporal seafloor motion from pressure sensor data, 
based on high-fidelity physics models, can be achieved in real time.
In that work, the authors created a \textit{digital twin}~\cite{DigitalTwinsNASEM} for tsunami early warning on the  Cascadia Subduction Zone (CSZ). The CSZ is a 1000~km long region spanning from Northern California to British Columbia (see~\Cref{fig:csz-candidate-sensors}), where %
paleoseismic
evidence suggests a magnitude 8.0--9.0 megathrust earthquake is overdue \cite{Atwater1995, Goldfinger2012}.

Community-driven initiatives and feasibility studies are currently laying the groundwork for the installation of offshore geophysical monitoring networks along the CSZ \cite{Schmidt2019,SZ4D2023}. 
While offshore initiatives like the NEPTUNE observatory \cite{Barnes2015} provide localized continuous observations, offshore sensor coverage remains sparse across the CSZ, limiting the ability to provide accurate tsunami forecasts. 
Japan's operational S-net system, which consists of 150 seafloor observation stations, 
demonstrates that such large-scale deployments are physically and technologically viable \cite{Aoi2020}. Yet, establishing an extensive underwater sensor network remains exceptionally expensive, raising a fundamental operational question: \textit{given a constrained budget, where should these sensors be placed to maximize the information gained from the sensors about the tsunami source?}

We address the challenge of sensor placement using the mathematical formalism of Bayesian optimal experimental design (OED)~\cite{huan2024optimal}. The work in~\cite{henneking2025bell} showed that sparse offshore observations can be used to infer tsunamigenic seafloor motion as part of a tsunami early warning system. Because this inverse problem is inherently ill-posed, a Bayesian framework is used to characterize uncertainty by combining prior physical knowledge with real-time data. Within this context, OED allows us to systematically identify the sensor locations that provide the maximum information gain about the model parameters, effectively minimizing the uncertainty in the inferred parameter field. 

Bayesian OED represents another ``outer loop'' added to the digital twin. The loops can be articulated as follows:
\begin{enumerate}
    \item \textbf{Inner loop (forward problem):} This loop consists of solving the forward PDEs that govern the underlying physical system. Through this, model parameters are mapped to system observables.
    \item \textbf{Middle loop (inverse problem):} This loop consists of inferring model parameters from observations. In general, the inverse problem is an optimization problem whose solution involves many forward problem solves (e.g., \cite{isaac2015scalable, seelinger2021high}).
    \item \textbf{Outer loop (OED):} This loop represents additional layers of optimization surrounding the inverse problem. 
    Solving the OED problem effectively requires the solution of many inverse problems (e.g., ~\cite{ghattas2021learning, alexanderian2021optimal}). 
\end{enumerate}

As a result, the computational expense of directly solving a Bayesian OED problem for a high-fidelity forward model and a high-dimensional parameter field is often insurmountable. To make large-scale Bayesian OED tractable, previous efforts have largely relied on exploiting low-rank structures intrinsic to the problem. Strategies such as randomized linear algebra \cite{koval2020optimal}, low-rank-approximation-based strategies \cite{alexanderian2014optimal, wu2023fast, alexanderian2016on}, and connections to the column subset selection problem \cite{eswar2026bayesian, diaz2025structured} have proven highly successful for parabolic or elliptic systems, where 
smoothing properties of 
PDE solution operators naturally lead to rapid spectral decay.

Unfortunately, these low-rank techniques are typically ineffective for problems governed by hyperbolic PDEs, such as the acoustic--gravity wave equations governing tsunami propagation. Hyperbolic systems suffer from the Kolmogorov $N$-width problem, which implies the non-existence of accurate low-dimensional linear subspace embeddings~\cite{greif2019kolmogorov}. For high-fidelity digital twins like the CSZ model in~\cite{henneking2025bell}, the underlying parameter space---here representing the spatiotemporal evolution of the seafloor---can easily exceed $\mathcal{O}(10^9)$ dimensions. Solving a Bayesian OED problem at this extreme scale has traditionally been considered computationally intractable. Indeed, the authors of~\cite{henneking2025bell} state that simply finding the mean of the posterior parameter field using standard methods would take 50 years on several hundred GPUs; solving the OED problem would be hundreds of thousands of times more expensive.

Recent advances have demonstrated that for linear time-invariant (LTI) dynamical systems, large-scale Bayesian inversion can be made extremely efficient. The Sherman--Morrison--Woodbury identity recasts the Bayesian inversion from the high-dimensional parameter space into the lower-dimensional data space~\cite{henneking2026goal}. Here, the dense data-space Hessian matrix ($\vb{K}$) can be explicitly constructed using highly optimized, FFT-based Hessian matrix-vector products for LTI systems\cite{henneking2025bell, venkat2025fft}. Using this methodology, we show that the previously intractable $\order{10^9}$-dimensional Bayesian OED problem can be turned into a purely combinatorial search over dense data-space Hessians. 

However, finding the optimal subset of sensors remains an NP-hard problem. While sequential OED can be posed as a dynamic programming problem \cite{huan2016sequential}, greedy algorithms are widely employed as a methodology for sensor subset selection~\cite{jagalur2021batch,wu2023fast,huan2024optimal}. 
Here, the main computational bottleneck is evaluating the OED objective (generally a log-determinant calculation) for many candidate sensors. A naive implementation of a greedy algorithm scales cubically with the number of selected sensors, quickly exhausting hardware memory limits and severely limiting the efficacy of the OED framework.

To bridge this final gap, this paper introduces a highly scalable, distributed-memory computational framework for D-optimal sensor selection. While demonstrated in the context of tsunami early warning, this algorithmic OED framework generalizes to the larger class of linear-Gaussian Bayesian inverse problems governed by LTI systems. 
To our knowledge, this framework is the first capable of 
carrying out PDE-constrained Bayesian OED over an $\mathcal{O}(10^9)$-dimensional parameter space. The main contributions of this work are as follows:

\begin{enumerate}
    \item \textbf{Algorithmic acceleration via Schur complement updates:} 
    We develop a parallel block Schur complement-based algorithm that completely avoids redundant dense matrix factorizations during the greedy candidate evaluation phase. This mathematical reformulation significantly reduces both the memory footprint and the computational time required to process the data-space Hessians.
    
    \item \textbf{Extreme-scale, performance-portable parallelization and pipelined I/O optimization:} 
    We co-design the Schur complement-based algorithm to target a massively parallel, multi-GPU architecture using PyTorch and MPI, ensuring performance portability across both AMD and NVIDIA GPU architectures. By utilizing strict zero-allocation in-place memory management, parallel POSIX I/O, and a double-buffered architecture with isolated compute streams, the framework completely overlaps I/O with computation. The implementation exhibits excellent weak and strong scalability on leadership-class supercomputers with varying architectures.
    
    \item \textbf{Application to digital twin for tsunami early warning:} 
    We demonstrate the efficacy of the proposed algorithms on an extreme-scale digital twin of the CSZ with over 1 billion degrees of freedom in the parameter field. We optimize a 175-sensor network from a pool of 600 candidates to minimize the uncertainties of the inferred parameters describing earthquake-induced seafloor motion. 
    To the best of our knowledge, this is the \textbf{largest-to-date solution of a PDE-constrained Bayesian OED problem} without using reduced-order or surrogate models. This paves the way for the cost-effective, real-world deployment of offshore early warning networks.
\end{enumerate}

The remainder of this paper is organized as follows. \Cref{sec:background} introduces the mathematical and computational formulations of Bayesian inverse problems for LTI systems. \Cref{sec:methods} details the greedy sensor selection algorithm, showing theoretical properties and highlighting parallel implementation details. \Cref{sec:single-gpu-results,sec:scaling-io-opt,sec:multi-gpu-results,sec:io-analysis} present single-GPU performance, parallel scaling methodology, multi-GPU scaling results, and pipelined I/O performance, respectively. Finally, \Cref{sec:oed-results} applies the framework to the CSZ digital twin to derive optimal sensor placements for tsunami early warning.

\section{Background}\label{sec:background}

The algorithms developed in this paper address Bayesian OED for inverse problems governed by LTI dynamical systems. In this section, we introduce Bayesian inverse problems in this setting and briefly outline a framework for efficiently computing their solutions. The methods described here are based on the work by Henneking et al.~\cite{henneking2026goal}.

\subsection{Bayesian inverse problems for LTI systems}
Mathematical models of wave propagation, transport, diffusion, and numerous other physical phenomena often take the form of LTI dynamical systems.
In many cases, the input parameters to these models, here denoted by $\vb{m}$, appear linearly in the governing equations (e.g., as boundary or volumetric sources).
At the same time, the model observables (e.g., observations of states at discrete sensor locations) are typically related to the PDE solution via a linear observation operator.
If these conditions are true, the \textit{parameter-to-observable} (p2o) map, denoted by $\vb{F} : \vb{m} \mapsto \vb{d}$, which maps the parameters $\vb{m}$ to the observables $\vb{d}$ via the PDE solution, also represents an LTI system.
In this case, both $\vb{F}$ and its adjoint $\vb{F}^*$ are time-shift-invariant; a temporal shift in the inputs (parameters) yields an identical shift in the outputs (observables). As a result, the discretized operators $\vb{F}$ and $\vb{F}^*$ are \textit{block-triangular Toeplitz} matrices~\cite{venkat2025fft}.

\begin{sloppypar}
The primary objective in Bayesian inversion is to characterize the \textit{posterior} probability distribution of the unknown model parameters $\paramvec$ given the (noisy) observational data $\obsvec$. Formulated in a discretized setting, Bayes' theorem states $\pipost(\paramvec | \obsvec)~\propto~\pilike(\obsvec | \paramvec) \piprior(\paramvec)$, i.e., the posterior distribution of $\paramvec$ is proportional to the product of the likelihood function and the prior distribution.
\end{sloppypar}

Assuming an additive, zero-mean Gaussian observational noise model with covariance matrix $\bfnoise$ and  a zero-mean Gaussian prior with covariance $\bfpriorcovm$, the posterior (see~\cite{stuart2010inverse}) is also a Gaussian, $\pipost \sim \mathcal{N}(\bfmmap, \bfpostcovm)$, where
\begin{align}
	\left( \ptomat^* \bfnoise^{-1} \ptomat + \bfpriorcovm^{-1} \right) \bfmmap &=
	\ptomat^* \bfnoise^{-1} \obsvec,\label{eq:map-point}\\
    \text{and} \quad \bfpostcovm \coloneqq \vb H^{-1} &= \left( \ptomat^* \bfnoise^{-1} \ptomat + \bfpriorcovm^{-1} \right)^{-1} .
	\label{eq:postcov}
\end{align}

\subsection{Efficient computational methods for LTI systems}\label{sec:computational}
While it is easy to give explicit expressions for the posterior mean and covariance in~\Cref{eq:map-point,eq:postcov}, actually solving the linear system in~\Cref{eq:map-point} is computationally challenging for high-dimensional problems. This effect is compounded for problems governed by hyperbolic systems, where the Hessian $\vb H$ lacks exploitable low-rank structure. However, efficient methods have recently been developed in~\cite{henneking2025bell} and~\cite{henneking2026goal} that enable real-time Bayesian inference for extreme-scale LTI systems. 

Within this real-time inversion framework, the inverse problem is partitioned into a set of computationally intensive offline precomputations and a rapid online phase for parameter inference. The offline phase primarily consists of computing adjoint PDE solutions, one per sensor location. After this setup phase, subsequent matrix-vector products with the discrete p2o map $\vb{F}$ are computed using efficient, FFT-based, GPU-accelerated algorithms that achieve $\order{10^5}\times$ 
speedups over traditional PDE solves~\cite{venkat2025fft, venkat2025mixed}.

A critical component of the inversion framework in~\cite{henneking2026goal} is the re-expression of the Hessian in the \textit{data space} using the Sherman--Morrison--Woodbury identity:
\bes
	\bfpostcovm
    = \left( \bfpriorcovm - \Gmat^* \Kmat^{-1} \Gmat \right) ,
\ees
where $\Kmat \coloneqq \bfnoise + \ptomat \Gmat^*$ and $\Gmat^* \coloneqq \bfpriorcovm \vb{F}^*$. The matrix $\vb{K}$ is termed the ``data-space Hessian,'' with dimension $N_dN_t \times N_dN_t$, where $N_d$ is the number of sensors, and $N_t$ is the number of timesteps. It can be explicitly formed as a dense matrix using the fast, FFT-based algorithms~\cite{henneking2026goal}. 

Thus, the dominant cost of the Bayesian inversion framework lies in computing the $\numdata$ adjoint PDE solutions that define $\vb{F}$. While each of these solves can take hours on hundreds of GPUs for large-scale problems, this is a one-time, offline setup cost. Moreover, once the adjoint PDE solutions have been computed for a superset of candidate sensor locations, inversions can be quickly evaluated using any smaller \textit{subset} of those locations. This enables us to ask the core question of this paper: \textit{given a fixed sensor budget, what are the optimal locations to select?}

In the remainder of the paper, we explore how this computational framework can be leveraged to facilitate Bayesian OED for extreme-scale systems. We then demonstrate how this framework enables the optimal design of seafloor sensor networks for the CSZ and allows for seamless integration of expert knowledge or practical cost constraints.

\section{Methods}\label{sec:methods}

We formalize the OED problem as an optimization problem that seeks to maximize the expected information gain (EIG) from data. The information gain is measured by the Kullback--Leibler (KL) divergence~\cite{kullback1951information} of the prior to the posterior.
For a linear inverse problem with Gaussian prior, Gaussian likelihood, and additive Gaussian noise, the problem of maximizing the EIG is equivalent to the problem of D-optimal experimental design~\cite{alexanderian2016on}. D-optimality seeks to minimize the determinant of the posterior covariance matrix. Another popular criterion is A-optimality, which minimizes the average posterior variance (the trace of the posterior covariance matrix)~\cite{alexanderian2014optimal, huan2013simulation}.

In this work, we focus on the D-optimal criterion primarily for two reasons. First, it possesses a clear information-theoretic interpretation as the KL divergence between the posterior and prior distributions. Second, as we will see in~\Cref{sec:submodularity}, the resulting objective function for OED is \textit{submodular}, making it an ideal choice for the greedy algorithm employed here. Thus, the remainder of the paper focuses on designing efficient algorithms for D-optimal experimental design for Bayesian inverse problems governed by LTI dynamical systems.

\subsection{D-optimal sensor subset selection}
Suppose we have a set $C$ of candidate sensors. Given a budget $B < |C|$, the goal of D-optimal sensor subset selection is to choose the set $S_{\text{opt}} \subset C$ of size $|S_{\text{opt}}| \leq B$ that minimizes the log-determinant of the posterior covariance. That is, we aim to find
\begin{align}
    S_{\text{opt}} 
    \coloneqq \argmin_{S \subset C, |S| \leq B} \Phi(S) 
    \coloneqq \log\det((\bfpostcov)_S),
    \label{eq:d-opt-obj}
\end{align}
where $(\bfpostcov)_S$ is the posterior covariance obtained using only information from the sensors in $S$.

\subsection{Notation}
Throughout the rest of the paper, we use the subscript $S$ to denote operators that contain only information from the sensors in $S$. For example, $\vb{F}_S$ and $\vb{G}_S$ are block-row subsampled versions of $\vb{F} \equiv \vb{F}_C$ and $\vb{G} \equiv \vb{G}_C$, respectively. Similarly, $\vb{K}_S$ is block-row and block-column subsampled from $\vb{K} \equiv \vb{K}_C$. For $\vb{F}$, $\vb{G}$, and $\vb{K}$, the subsampled operators can be formed by directly subsampling the respective full operators along the appropriate dimensions. However, an operator such as $(\bfpostcov)_S$, which involves the \textit{inverse} of subsampled operators, must be formed from its expression in \Cref{sec:computational} using the subsampled operators $\vb{F}_S$, $\vb{G}_S$, and $\vb{K}_S$. Importantly, $(\bfpostcov)_S$ is \textit{not} equal to a subsampled version of the corresponding full operator $\bfpostcov \equiv (\bfpostcov)_C$.

\subsection{Computable D-optimal objective function}
The D-optimal objective function in the form of \Cref{eq:d-opt-obj} is not directly computable for high-dimensional parameter fields, where forming the covariance matrix $\bfpostcov$ is intractable.
Instead, we reformulate the D-optimal objective function in the (lower-dimensional) data space by using the matrix determinant lemma~\cite{harville1997matrix}:
\begin{align*}
\det((\bfpostcov)_S) = &\det\qty(\bfpriorcov - \vb{G}_S^*\vb{K}_S^{-1}\vb{G}_S) \\
 = &\det(\bfpriorcov)\det(\vb{K}_S^{-1})\det\qty((\bfnoise)_S).
\end{align*}

Thus, the D-optimality objective reduces to
\begin{align}
    \Phi(S) = -\log\det(\vb{K}_S) + \log\det((\bfnoise)_S).\label{eq:d-opt-obj-computable}
\end{align}
Moreover, with the added assumption of isotropic noise (i.e., $\bfnoise = \gamma^2 \vb{I}$), the D-optimality objective becomes
\begin{align}
    \Phi(S) = -\log\det(\vb{K}_S).\label{eq:d-opt-obj-computable-id-noise}
\end{align}

Given the full matrices $\vb{K}$ and $\bfnoise$, evaluating the D-optimal objective becomes computationally efficient when the data space is of a lower dimension than the parameter space. For a given sensor subset $S \subset C$, these matrices can be directly subsampled to form $\vb{K}_S$ and $(\bfnoise)_S$, respectively. The log determinants can be computed via Cholesky factorization.

Note that an equivalent formulation of the D-optimal design problem is given by defining $\Tilde{\Phi} \coloneqq - \Phi$ and solving
\begin{align}
    S_{\text{opt}} 
    \coloneqq \argmax_{S \subset C, |S| \le B} \tilde{\Phi}(S)
    \label{eq:d-opt-obj-mod}
\end{align}
instead. This is the approach used in the remainder of the paper.

\subsection{Greedy algorithm for computing D-optimal design}

While the computable forms of the objective in~\Cref{eq:d-opt-obj-computable,eq:d-opt-obj-computable-id-noise} can be efficiently evaluated, the combinatorial optimization problem in~\Cref{eq:d-opt-obj-mod} is NP-hard~\cite{welch1982algorithmic, civril2009selecting}. We employ a greedy algorithm for approximating the solution to this optimization problem. Given a budget $B$ and a matrix $\vb{K}$ for a candidate sensor set $C$, the greedy algorithm builds the (approximate) optimal sensor set $S_{\text{opt}}$ iteratively. Let $S_{\text{opt}}^k$ be the greedy-optimal sensor set at iteration $k$. At iteration $k+1$, sensor $s$ is added to $S_{\text{opt}}^k$ if it provides the maximum information gain to the current sensor selection, i.e., if it maximizes $\Tilde{\Phi}\qty(S_{\text{opt}}^k \cup \qty{s}) - \Tilde{\Phi}\qty(S_{\text{opt}}^k)$ among all sensors in $C\setminus S_{\text{opt}}^k$. 

Evaluating $\Tilde{\Phi}\qty(S_{\text{opt}}^k \cup \qty{s})$ for each of the candidate sensors $s\in C\setminus S_{\text{opt}}^k$ involves calculating the log-determinants of the ``test matrices'' $\vb{K}_{S_{\text{opt}}^k \cup \qty{s}}$. The determinants are evaluated using Cholesky factorization. However, instead of refactorizing the test matrix from scratch for each candidate sensor, a block Cholesky update algorithm is used. The algorithm, assuming isotropic noise, is given in~\Cref{alg:opt-sens-select-greedy-seq}. It can easily be modified for the anisotropic noise case.

\subsection{Algorithmic complexity}\label{sec:complexity}

The computational efficiency of the greedy selection process is dependent on how the information gain is evaluated for each candidate. After $k$ iterations of the algorithm, the $(k+1)$-th iteration evaluates the $|C| - k$ remaining candidate sensors. For each candidate $s$, the test matrix $\vb{K}_{S_{\text{opt}}^k \cup \qty{s}}$ is formed by augmenting $\vb{K}_{S_{\text{opt}}^k}$ with the dense block rows and columns of $\vb{K}$ corresponding to $s$; the resulting test matrix has size $(k+1)N_t \times (k+1)N_t$. A naive implementation that evaluates the objective function by refactorizing the test matrix from scratch requires $\mathcal{O}(k^3 N_t^3)$ floating-point operations per candidate. Summed over the full selection budget $B$, this approach scales with a cumulative complexity of $\mathcal{O}(|C| B^4 N_t^3)$.

By leveraging the block Cholesky update and the Schur complement, the optimized formulation strictly avoids redundant factorizations. Because the Cholesky factor $\vb{L}_S$ from iteration $k-1$ is reused, evaluating a new candidate only requires solving a lower-triangular system via forward substitution. This step dominates the per-candidate cost, requiring only $\mathcal{O}(k^2 N_t^3)$ operations. Consequently, the block update strategy reduces the total algorithmic complexity across the entire selection process to $\mathcal{O}(|C| B^3 N_t^3)$, effectively lowering the computational cost by a factor equal to the number of budgeted sensors---which can be very large.

\subsection{Multi-GPU implementation and memory optimizations}\label{sec:parallel-implementation}

As the evaluation of candidate sensors is embarrassingly parallel, \Cref{alg:opt-sens-select-greedy-seq} can be easily extended to incorporate distributed memory parallelism. First, the candidate sensor evaluations are distributed among the processors participating in the computation. At each iteration, once the next sensor has been found, the optimal score value $d_{\max}$ and the integer index of the selected sensor are communicated to all processors. Then, the processors update their local copies of $\vb{L}_S$ accordingly. This strategy, while incurring some redundant computation, avoids the expensive communication of the $\vb{L}_S$ matrix over the network. In addition, all linear algebra operations are performed strictly in-place to minimize the GPU memory footprint, as this is the primary factor constraining the maximum sensor budget $B$.

Each candidate evaluation in~\Cref{alg:opt-sens-select-greedy-seq} consists of two main phases: reading the required matrix blocks from the full $\vb{K}$ matrix and computing the objective function to get the score value $d_s$ for candidate sensor $s$. As the processors independently evaluate candidates, they execute thousands of scattered read requests to fetch the non-contiguous $N_t \times N_t$ blocks required to build the test columns $\vb{K}_{S,s}$. At iteration $k$, for each candidate $s$, $k$ blocks must be fetched from memory. To maximize network bandwidth and avoid the metadata lock contention that traditionally throttles collective access on parallel file systems~\cite{thakur1999implementing,latham2004impact,latham2013parallel}, the dense covariance matrix $\vb{K}$ is stored as a 2D-chunked HDF5 dataset and accessed via independent parallel POSIX I/O~\cite{folk2011overview,howison2010tuning,latham2012case}. This implementation was found to deliver better performance than independent or collective MPI-IO thanks to decreased latencies on the scattered access pattern.

Crucially, the implementation was designed to maximize overlap of I/O with computations: after the CPU retrieves the data for one candidate from the file system, it initiates an asynchronous host-to-device memory transfer to move the data to the GPU. The GPU then performs the necessary objective function evaluation for that candidate. In the meantime, the CPU issues another set of read requests to fetch the next candidate from the file, using a double-buffered framework. Furthermore, this system is coupled with isolated CUDA/HIP streams to create a fully overlapped, asynchronous execution pipeline: while the CPU thread executes a blocking POSIX read of candidate $s+1$ from the file system into pinned host memory, a dedicated GPU stream concurrently dispatches the non-blocking host-to-device (H2D) PCIe transfer and the dense in-place objective function calculations for candidate $s$. Fine-grained CUDA/HIP events synchronize the pipeline stages to prevent race conditions, ensuring that I/O and memory transfers are entirely hidden behind the execution of the GPU computations.

Following this inner loop of candidate evaluations, an \texttt{MPI\_Allreduce} operation is used to synchronize the ranks and determine the global maximum $d_{\max}$ and corresponding sensor index $s^*_{\text{global}}$. 
Because the dense matrix components were intentionally discarded to avoid broadcasting large blocks over the network, a recomputation phase is inserted immediately before the global state update. Every rank independently re-reads the data for $s^*_{\text{global}}$ and re-executes the forward substitution and Schur complement specifically for the global maximizer. This synchronous recomputation enables all workers to execute the block Cholesky update in parallel, ensuring that each rank maintains an identical, up-to-date copy of the global factor $\vb{L}_S$ without overwhelming the interconnect bandwidth.

To further accelerate the GPU execution, the algorithm relies on a globally pre-allocated Cholesky factor. Since the final sensor budget $B$ is known a priori, the absolute maximum memory footprint of the lower-triangular factor $\vb{L}_S$ is bounded by a $(B N_t) \times (B N_t)$ block matrix. This buffer is allocated on the accelerator once during initialization. During iteration $k$, the parallel workers operate exclusively on a $k N_t \times k N_t$ tensor view of this contiguous block. This zero-allocation inner loop prevents GPU memory fragmentation and entirely eliminates the latency of dynamic memory allocation during the computationally intensive candidate evaluation phase.

\begin{algorithm}
\caption{Sequential greedy algorithm for block D-optimal sensor selection. This algorithm assumes isotropic noise, but can be easily modified to handle extensions. $\vb{K}$ is the dense data-space Hessian, $C$ is the set of candidate sensors (corresponding to block-rows and columns of $\vb{K}$), and $B$ is the sensor selection budget (i.e., number of sensors to be selected).} \label{alg:opt-sens-select-greedy-seq}
\begin{algorithmic}[1]
\Procedure{select\_sensors}{$\vb{K}$, $C$, $B$}
\State $S \gets \emptyset$
\State $\vb{L}_S \gets [\,]$ \Comment{\textcolor{blue}{Initialize empty global Cholesky factor}}
\For{$k = 1\dots B$}
    \State $d_{\max}, s^* \gets -\infty, \text{null}$
    \For{$s\in C\setminus S$}
        \If{$k = 1$}
            \State $\vb{M}_s \gets \vb{K}_{s,s}$ 
        \Else \Comment{\textcolor{blue}{Triangular solve + Schur complement}}
            \State Solve $\vb{L}_S \vb{Y}_s = \vb{K}_{S,s}$ for $\vb{Y}_s$ 
            \State $\vb{M}_s \gets \vb{K}_{s,s} - \vb{Y}_s^\top \vb{Y}_s$ 
        \EndIf
        
        \State $\vb{L}_{M,s} \gets \text{Cholesky}(\vb{M}_s)$ 
        \State $d_s \gets 2\cdot \text{Sum}\qty(\log \qty(\text{diag}(\vb{L}_{M,s})))$ \Comment{\textcolor{blue}{Log-det.}}
        
        \If{$d_s > d_{\max}$}\Comment{\textcolor{blue}{Update best candidate}}
            \State $d_{\max}, s^*,  \vb{L}_{M^*} \gets d_s, s, \vb{L}_{M,s}$
            \If{$k > 1$}
                \State $\vb{Y}_{s^*} \gets \vb{Y}_s$ 
            \EndIf
        \EndIf
    \EndFor
    \State $S \gets S \cup \qty{s^*}$ \Comment{\textcolor{blue}{Add new sensor to optimal set}}
    \If{$k = 1$}
        \State $\vb{L}_S \gets \vb{L}_{M^*}$ 
    \Else
        \State $\vb{L}_S \gets \qty[\begin{smallmatrix} \vb{L}_S & \vb{0} \\ \vb{Y}_{s^*}^\top & \vb{L}_{M^*} \end{smallmatrix}]$ \Comment{\textcolor{blue}{Block update}}
    \EndIf
\EndFor
\State \Return $S$
\EndProcedure
\end{algorithmic}
\end{algorithm}

\subsection{Modifications to the D-optimality objective}
The objective function in~\Cref{eq:d-opt-obj-computable-id-noise} considers only the EIG. This objective function can be modified to account for sensor cost and to apply a nonuniform weighting to the posterior covariance $\bfpostcov$.

\subsubsection{Sensor cost considerations}
The cost of placing seafloor sensors can vary based on a number of factors, including depth, access rights, and ease of connectivity to the rest of the network. Such variable cost factors can be incorporated into the OED objective by using a weighted noise covariance $\tilde{\vb{\Gamma}}_{\!\text{noise}} = \vb{W}_{\! c}\bfnoise$. Here, $\vb{W}_{\! c}$ is a matrix that determines the cost of placing each sensor. If the sensor cost is only dependent on properties (e.g., location) of the individual sensor, $\vb{W}_{\! c}$ is a diagonal matrix. However, a more complicated, cross-correlated cost structure can also be used. By using the matrix $\vb{K}$ created with the weighted noise covariance, the OED problem can incorporate sensor placement costs without modifying the sensor selection algorithm. The scale of the weight matrix $\vb{W}_{\! c}$ relative to $\vb{F}\vb{G}^*$ and $\bfnoise$ is a hyperparameter that can be tuned on a problem-specific basis.

\subsubsection{Nonuniform weighting of uncertainties}\label{sec:masking}
In practical applications, the posterior uncertainty field might not be uniformly important over the entire spatiotemporal parameter field. For example, we may want to reduce parameter uncertainties more in specific parts of the domain, and we may not care at all about the parameter uncertainties in other parts of the domain. Similarly, we may care more about reducing parameter uncertainties in a specific time interval. These considerations can easily be incorporated into the framework described above by modifying the full $\vb{K}$ matrix once before performing the optimal sensor selection. Recall that $\vb{K} = \bfnoise + \vb{F}\bfpriorcov\vb{F}^*$. We define the modified matrix $\tilde{\vb{K}} \coloneqq \bfnoise + \vb{F}\vb{W}_{\! m}\bfpriorcov\vb{W}_{\! m}^*\vb{F}^*$, where $\vb{W}_{\! m}$ is a diagonal weighting matrix that applies the nonuniform weighting to the spatiotemporal parameter field. Then, we can run the sensor selection~\Cref{alg:opt-sens-select-greedy-seq} to determine the optimal sensors with respect to this weighted measure of uncertainties.

\subsection{Submodularity of the D-optimal objective}\label{sec:submodularity}
As mentioned previously, the full combinatorial optimization problem in~\Cref{eq:d-opt-obj} is NP-hard. While we use a greedy algorithm to obtain an approximate solution, it is important to understand the theoretical bounds associated with this approximation.
For this purpose, we first define the notion of a \textit{submodular} function. 
\begin{definition}[Submodularity]
A set function $f: 2^V \to \mathbb{R}$ is submodular if for all subsets $X \subseteq Y \subset V$ and any element $e \in V \setminus Y$, we have:
\[
    f(X \cup \{e\}) - f(X) \ge f(Y \cup \{e\}) - f(Y) .
\]
This property formalizes the concept of ``diminishing returns.''
\end{definition}

For a symmetric positive definite (SPD) matrix $\vb{K}$, the submodularity of the log-determinant of its principal submatrices is a well-established result that can be shown via information-theoretic methods relating the objective to Shannon entropy of a multivariate Gaussian, which is fundamentally submodular~\cite{cover2006elements}. The result can also be derived from standard linear algebraic methods using the properties of Schur complements~\cite{ko1995exact}.

Furthermore, foundational results for submodular maximization~\cite{nemhauser1978analysis,alexanderian2026submodularity} ascertain that the optimal sensor set obtained from our greedy algorithm is at least $(1-1/e) \approx 63\%$ as good as the globally optimal sensor set.

While the submodularity of the D-optimal objective function naturally suggests the use of the lazy greedy (or accelerated greedy) algorithm to reduce the total number of candidate evaluations (e.g.~\cite{minoux1978accelerated}), such approaches are structurally incompatible with the large-scale distributed architectures we target. The lazy greedy algorithm reduces the total number of candidate evaluations by maintaining a priority queue of stale upper bounds on marginal gains, re-evaluating a candidate only when its bound remains the global maximum. This approach trades evaluations for sequential dependencies: within each iteration, candidates must be evaluated in priority order. In a distributed setting, this serialization prevents assigning disjoint candidate subsets to independent ranks. The tradeoff of lazy greedy is unfavorable in our regime. 

As shown in~\Cref{sec:results}, the per-candidate GPU compute time is $\order{10^{-1}}\!\text{ s}$ while the \texttt{MPI\_Allreduce} used to determine the global winner takes $\order{10^{-4}}\!\text{ s}$, so communication is not the bottleneck. Lazy greedy is beneficial when individual candidate evaluations are expensive. However, the efficiency of the Schur complement updates puts us in the opposite regime: evaluations are fast, parallelism is abundant, and the sequential queue structure of lazy greedy would negate the scaling advantage of concurrent candidate evaluations. %
Lazy greedy could be an effective alternative if available GPU resources are limited.

\section{Results}\label{sec:results}
The multi-GPU version of~\Cref{alg:opt-sens-select-greedy-seq} was implemented using PyTorch and mpi4py to enable performance portability and ease of use. The dense matrix $\vb{K}$ is stored on disk in double precision (\texttt{float64}), as it is used elsewhere in the Bayesian inversion framework where stable inversion from noisy data precludes the use of lower precision calculations. However, all dense linear algebra computations within the candidate evaluation loop are performed in single precision (\texttt{float32}). Because the exact scalar values of the objective function are utilized solely to determine the relative ordering of candidate configurations during the greedy selection phase, single precision provides sufficient accuracy while halving the active memory footprint and providing significantly higher GPU throughput. The single-precision computations were also compared to double-precision versions, and the obtained sensor selections agree exactly for a given sensor budget. %
\iftoggle{cameraready}
{
The code is available open-source in the AutonomousOED package (\url{https://github.com/s769/AutonomousOED}).
}{
}

To validate the theoretical algorithmic complexity and systems-level optimizations detailed in the previous sections, we developed a suite of application-oriented computational benchmarks. These benchmarks evaluate both the single-GPU performance of the proposed block Schur complement update method and the distributed-memory parallel scalability of the fully-pipelined sensor selection framework. The benchmarks are designed to evaluate the performance of the OED implementation at the full scale of the tsunami digital twin application as well as to determine its scalability to even larger problem sizes. Comprehensive details regarding hardware dependencies, software environment versions (e.g., PyTorch, mpi4py, HDF5), and environment variables required to reproduce these experiments are provided in the AD/AE appendices.

\subsection{Single-GPU performance results}\label{sec:single-gpu-results}
To quantify the single-GPU performance of the algorithmic design and implementation, a computational study was conducted across five GPU architectures: an NVIDIA A100 (80~GB), an AMD MI250X (64~GB), an NVIDIA GH200 Superchip (96~GB), an NVIDIA GB200 Superchip (192~GB), and an AMD MI300X (192~GB). The study considers two implementations of the objective function evaluation step:
\begin{enumerate}
    \item \textbf{Naive formulation:}\footnote{``Naive'' here refers only to the method of fully refactorizing the test matrix for each candidate sensor. The OED algorithm itself is still based on the efficient data-space formulation; without this formulation, the OED problem would be entirely intractable (see~\Cref{sec:introduction}).} Computes the objective by refactorizing the full augmented test matrix from scratch for each candidate.
    \item \textbf{Schur formulation:} The proposed method, leveraging the block Schur complement update to avoid redundant factorizations.
\end{enumerate}

Both implementations employ strict in-place memory operations to minimize the GPU memory footprints, which increases the possible budget size $B$. Intermediate tensors are pre-allocated outside the inner loop. The default PyTorch backend is used for the linear algebra operations, which uses an internal heuristic to determine whether cuSOLVER/hipSOLVER or MAGMA~\cite{abdelfattah2017fast} is called at runtime.

For this benchmark, the number of timesteps (i.e., the observations per sensor location) is fixed to $N_t = 420$, selected based on a realistic sampling frequency of 1~Hz over a 7-minute inversion time window~\cite{henneking2026real}. The computational cost per sensor selection iteration also depends on the current iterate $k$, which defines the number of already-selected sensors, and the total budget $B$. In the benchmark, $k$ is swept from 10 sensors to a maximum budget of 600 in increments of 10. For each value of $k$, we measure the peak GPU memory allocated and the average compute time per candidate over 10 independent runs. If an algorithm variant exceeds the device's specific memory limit for a given $k$, it is marked as an out-of-memory (OOM) failure and excluded from subsequent, larger budget evaluations. CUDA/HIP events are used to measure the GPU execution times, averaging over 10 runs after a warm-up period (we note that the observed variance between the independent runs is negligible).

\begin{figure*}[htbp]
    \centering
    \begin{subfigure}[b]{0.48\textwidth}
        \centering
        \includegraphics[width=\textwidth,trim={0 10pt 0 0}]
        {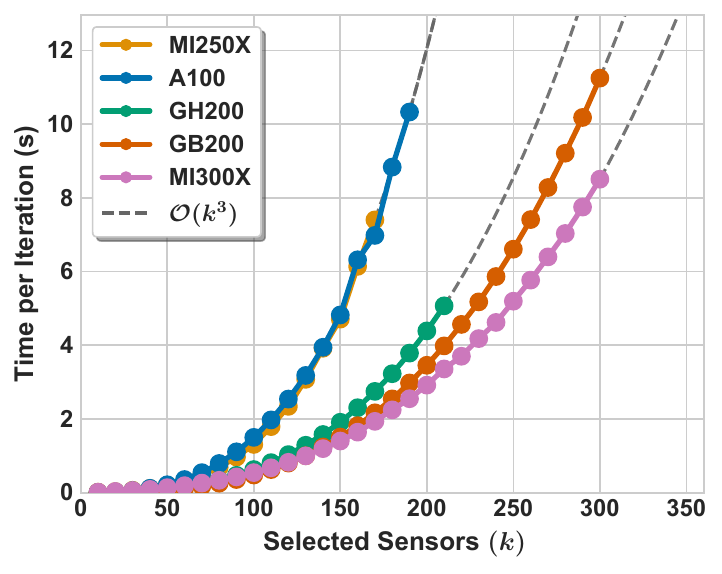}
        \caption{Naive Formulation ($\mathcal{O}(k^3)$)}
        \label{fig:naive-time}
    \end{subfigure}
    \hfill
    \begin{subfigure}[b]{0.48\textwidth}
        \centering
        \includegraphics[width=\textwidth,trim={0 10pt 0 0}]
        {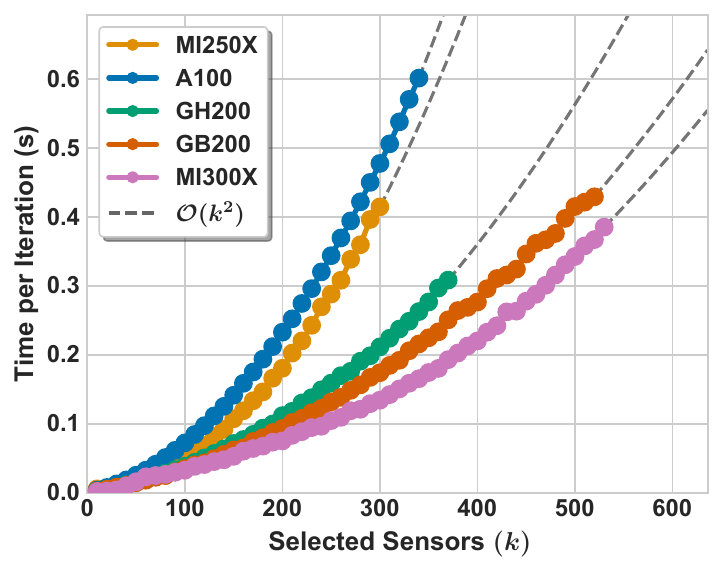}
        \caption{Schur Formulation ($\mathcal{O}(k^2)$)}
        \label{fig:schur-time}
    \end{subfigure}
    \caption{Single-GPU execution times per candidate evaluated across the five accelerator architectures. (a) The naive formulation exhibits $\mathcal{O}(k^3)$ polynomial scaling. (b) The Schur complement formulation reduces the complexity to $\mathcal{O}(k^2)$, significantly reducing the compute time for each candidate evaluation and enabling practical scaling to massive candidate pools. Dashed gray lines represent theoretical asymptotic behavior. The maximum number of selected sensors for each GPU architecture is determined by its memory limit.}
    \label{fig:ablation-time}
\end{figure*}

 The benchmark results are shown in \Cref{fig:ablation-time}. The plots illustrate that the Schur complement algorithm (plot b) is several orders of magnitude faster than the naive algorithm (plot a). Both the naive and Schur implementations follow the expected asymptotic behaviors estimated in~\Cref{sec:complexity}. Both algorithms have a memory complexity scaling with $\order{k^2}$, but the Schur algorithm's overall memory footprint remains much smaller. The relatively larger storage requirement of the naive algorithm results in a decreased maximum budget size. Using the Schur formulation, the maximum budget is 300 on the 64~GB MI250X (single GCD), 340 on the 80~GB A100, 370 on the 96~GB GH200, 520 on the 192~GB GB200, and 530 on the 192~GB MI300X. We use the more memory-efficient and higher-performing Schur formulation for testing the scalability of our OED implementation on the target machines: Perlmutter and Frontier.

\subsection{Parallel scaling methodology and I/O optimization}\label{sec:scaling-io-opt}

We assess the distributed-memory performance of our OED algorithm with scaling studies across two distinct supercomputing facilities: 
NERSC's Perlmutter and OLCF's Frontier. 
Perlmutter's GPU partition features an AMD EPYC 7763 64-core CPU and 4 NVIDIA A100 GPUs per node, backed by a 35~PB all-NVMe Lustre scratch file system. 
Frontier, the first exascale system, features an AMD EPYC 7713 64-core CPU and 4 AMD MI250X GPUs (8 Graphics Compute Dies, or GCDs) per node, backed by the Orion file system based on Lustre and HPE ClusterStor. In all tests, one MPI rank is used per GPU device/GCD.

In our strong and weak parallel scalability studies, the number of timesteps $N_t$ is kept fixed at 420. The global problem size for strong scaling uses 8,192 candidate sensors, and the local problem size for weak scaling uses 256 candidate sensors per GPU. 
The scalability tests measure the algorithm's performance at the $k = 175$-th iterate of the sensor selection.\footnote{Computing scaling results over the full OED algorithm would require long compute times. While the sensor selection times vary for each iterate, and generally grow with $k$ (see \Cref{sec:complexity}), we note that similar weak and strong scaling results were observed for other iterates, $k < 175$ and $k > 175$.}
Scaling studies were conducted with up to 1,024 GPUs on Perlmutter and Frontier (up to 2,048 GCDs).

With these scaling configurations, the largest problem sizes for weak scalability would require storing candidate matrices %
that can exceed practical limitations of filesystem access on shared computing facilities. However, for the purpose of testing the scalability of the OED algorithm, we can instead use a smaller candidate matrix, based on just 600 candidate sensors, and simulate larger numbers of candidates by using a modulo indexing system: candidate indices are mapped modulo 600 to the HDF5 dataset. To prevent unwanted filesystem caching effects, the global list of candidates is randomized before being distributed to the individual GPUs. The full $252{,}000\times 252{,}000$ candidate matrix $\vb{K}$ is stored in double precision as a 508~GB chunked HDF5 2D dataset.

To maximize I/O throughput, cross-facility file stripe sizes were set according to system guidelines. On Perlmutter, the HDF5 file was explicitly restriped across 128 Lustre OSTs with 2~MB extents. On Frontier's Orion file system, manual striping was intentionally omitted, because the dataset fell below the 512~GB manual-striping threshold. Thus, the default Progressive File Layout (PFL) was utilized, allowing the OLCF system wrapper to automatically manage optimal data distribution. Additionally, the internal HDF5 chunk cache and file locking mechanisms were disabled (\texttt{rdcc\_nbytes=0}, and \texttt{export~HDF5\_USE\_FILE\_LOCKING=FALSE}). This further increased performance by reducing CPU memory copies and unwanted blocks or synchronizations.

While the parallel file systems on Perlmutter and Frontier are capable of multi-TB/s peak theoretical bandwidths for large, contiguous collective operations, our Bayesian OED algorithm fundamentally relies on dynamically fetching small, non-contiguous blocks. Under this highly scattered, random-access POSIX workload, the file system becomes bound by metadata and seek latency, not data transfer limits. As a result, we aim to hide latencies rather than saturate the peak I/O bandwidth. 
Our current implementation does not use MPI RMA or NVMe burst buffers, which could further reduce I/O latency. However, because our fully-overlapped pipeline already hides I/O behind computation, the implementation is no longer I/O-bound.

Execution timers were recorded locally on all MPI ranks using \texttt{MPI\_Wtime}. %
The reported runtimes correspond to the maximum total wall time for all candidate evaluations across all ranks. Each scaling test was performed several times at different periods throughout the day to account for variations in filesystem performance; the observed variance between the independent runs was negligible. The communication times of the \texttt{MPI\_Allreduce} (for determining the global maximizer or best candidate) were also measured; these are negligible compared to the overall computation and I/O times.

\subsection{Multi-GPU scalability results}\label{sec:multi-gpu-results}
Strong scaling reflects the main operational goal of the framework: minimizing the time-to-solution for a fixed candidate pool. Solving the OED problem is a ``one-time'' offline computation, but the particular OED formulation in practice depends on external factors such as sensor deployment cost models, expectations about potential earthquake scenarios, and other prior knowledge that may be incorporated in the OED problem formulation (see~\Cref{sec:masking}). Decreasing time-to-solution will thus enable rapid OED computations for different configurations of these external parameters. Decreasing the time for objective function evaluations may also enable alternative optimization algorithms that require more candidate evaluations than our greedy selection approach does. 

To evaluate the framework's ability to maintain parallel efficiency as the problem size grows proportionally with the hardware, a weak scaling study was also conducted. The workload was fixed at 256 candidate evaluations per MPI rank, scaling the total number of candidate evaluations up to 262,144 on Perlmutter and 524,288 on Frontier.

\begin{figure*}[htbp]
    \centering
    \begin{subfigure}[b]{0.48\textwidth}
        \centering
        \includegraphics[width=\textwidth,trim={5pt 5pt 0 0},clip]
        {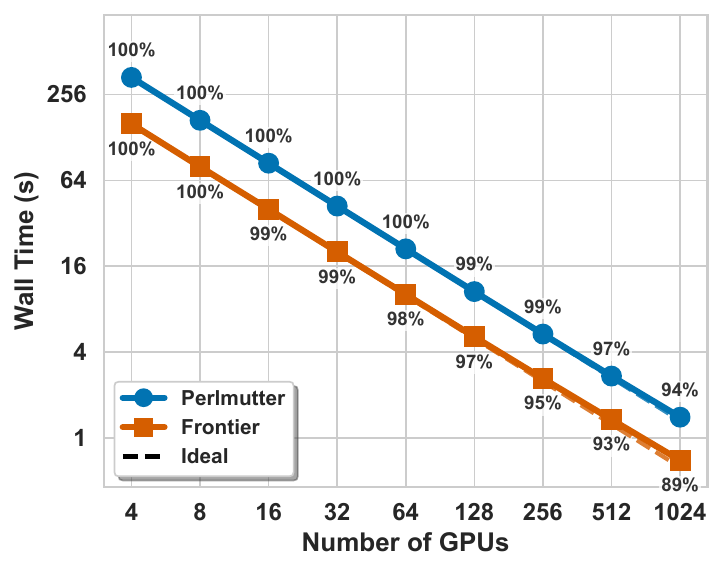}
        \caption{Strong Scaling Performance}
        \label{fig:strong_dual}
    \end{subfigure}
    \hfill
    \begin{subfigure}[b]{0.48\textwidth}
        \centering
        \includegraphics[width=\textwidth,trim={5pt 5pt 0 0},clip]
        {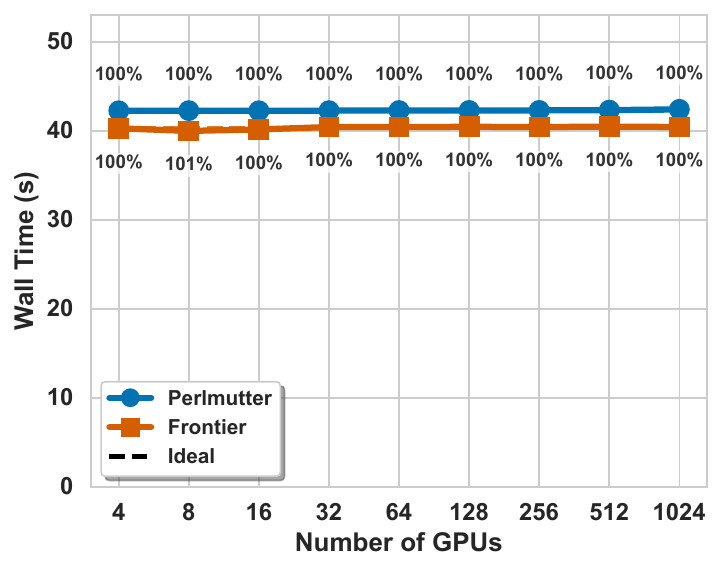}
        \caption{Weak Scaling Performance}
        \label{fig:weak_dual}
    \end{subfigure}
    \caption{Parallel (a) strong and (b) weak scaling performance of the fully-pipelined OED algorithm on NERSC's Perlmutter and OLCF's Frontier supercomputers. Observed wall times are plotted alongside ideal (linear) scaling baselines. The scaling efficiency percentage relative to the initial single-node baseline is annotated beside each marker, demonstrating robust parallel scalability across a $256\times$ increase in GPUs. Note that each GPU on Frontier consists of 2 GCDs.}
    \label{fig:scaling-results}
\end{figure*}

\Cref{fig:scaling-results} shows the strong and weak scalability results. We observe near-ideal weak and strong scaling over a $256\times$ processor increase on both Perlmutter and Frontier. While the computation phase of the algorithm is embarrassingly parallel, it is the pipelined execution scheme masking I/O behind compute that allows the multi-GPU implementation to reach this high parallel efficiency. Initial implementations involving sequential I/O-compute phases did not exhibit this level of scalability. Note that while the $x$-axis on~\Cref{fig:scaling-results} reports number of GPUs, each GPU on Frontier comprises 2 GCDs; the effective number of devices on Frontier is doubled.

\subsection{Pipelined I/O performance analysis}\label{sec:io-analysis}
\begin{figure*}[htbp]
    \centering
   \includegraphics[width=\textwidth,trim={25pt 40pt 0 0},clip]
   {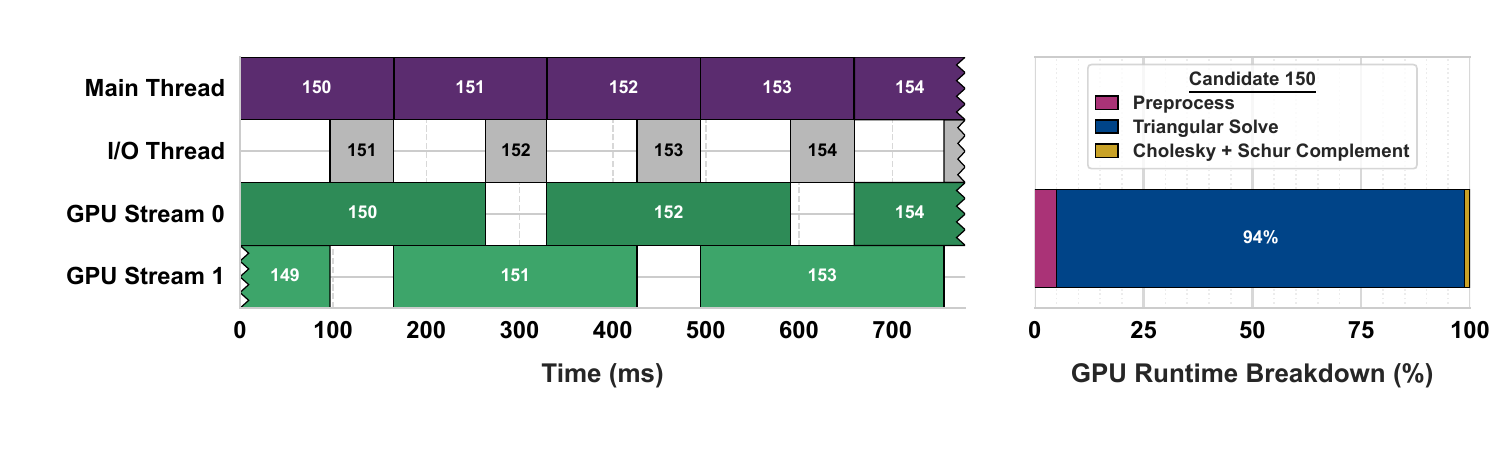}
   \caption{Execution timeline trace on one A100 GPU during a 16-GPU weak-scaling run on NERSC's Perlmutter at the \mbox{$k=175$-th} sensor-selection iterate (left), and GPU runtime breakdown for one candidate evaluation (right) during this iterate. Candidate I/O is overlapped with computation on two GPU streams, the main host thread remains occupied enqueuing GPU compute kernels, and the GPU remains busy throughout the execution timeline. The right panel shows that the triangular-solve stage of the candidate evaluation accounts for the vast majority of the GPU runtime.}
    \label{fig:io-timeline}
\end{figure*}

To assess the performance and effectiveness of the fully-pipelined OED algorithm, a detailed execution profile was conducted. The left panel of \Cref{fig:io-timeline} shows a portion of the execution timeline on one A100 GPU during a weak-scaling run on 16 A100 GPUs of Perlmutter during the \mbox{$k=175$-th} sensor-selection iterate (similar results were observed on other systems and for other iterates). In this plot, the ``main thread'' represents the CPU host thread that is enqueuing operations (kernels) to the GPU. The ``I/O'' thread is responsible for fetching the necessary data blocks from the HDF5 file stored on disk containing the matrix $\vb{K}$. ``GPU Stream 0/1'' represent the GPU-local computations. 
The right panel of \Cref{fig:io-timeline} shows a breakdown of the GPU compute time for a single candidate evaluation. The primary computational cost stems from computing the triangular solve using \texttt{torch.linalg.solve\_triangular()} (line 10 of \Cref{alg:opt-sens-select-greedy-seq}). 

The performance analysis shows that I/O is effectively hidden behind GPU computation, and the GPU remains busy throughout the execution. The throughput at which candidates are processed is instead governed by the rate at which compute kernels are submitted by the host and executed on the device. A detailed profile analysis reveals that each call to \texttt{torch.linalg.solve\_triangular()} expands into many internally launched GPU BLAS kernels (e.g., matrix multiply and triangular update operations). As a result, the main thread remains occupied until all of these internal kernels have been submitted to the GPU. Thus, even if the next candidate data has been fetched, GPU computation can only begin once the main thread has finished submitting kernels for the previous candidate. Experiments that used more than two buffers for I/O allowed candidate data to be fetched earlier, but did not increase the overall candidate throughput. Adding additional GPU streams increased compute concurrency, but also increased the duration of each candidate evaluation, resulting in no net improvement in candidate throughput. Finally, increasing the number of host kernel submission threads introduced substantial overhead that again resulted in no net improvement in candidate throughput. 
These results indicate that the current double-buffered implementation provides the optimal balance between I/O-compute overlap, compute concurrency, and per-candidate evaluation time.
This implementation achieves a speedup of \mbox{$\sim\! 3.5\times$} over a non-pipelined, single-buffered implementation.

\subsection{Optimal sensor placement for Cascadia digital twin}\label{sec:oed-results}
We applied our optimal sensor selection algorithm to a digital twin for tsunami early warning in the Cascadia Subduction Zone (CSZ), using the framework proposed in the Gordon Bell Prize-winning work~\cite{henneking2025bell}. In that work, the authors presented a digital twin that can infer the earthquake-induced seafloor motion %
from seafloor pressure observations %
in real time, and use the inferred field to forecast tsunami wave heights. This was demonstrated for a realistic Cascadia earthquake rupture scenario~\cite{glehman2025partial} using synthetic data from 600 hypothesized seafloor sensor locations across the CSZ. %

\begin{figure}[htbp]
    \centering
    \includegraphics[width=\columnwidth,trim={0 10pt 0 0},clip]
    {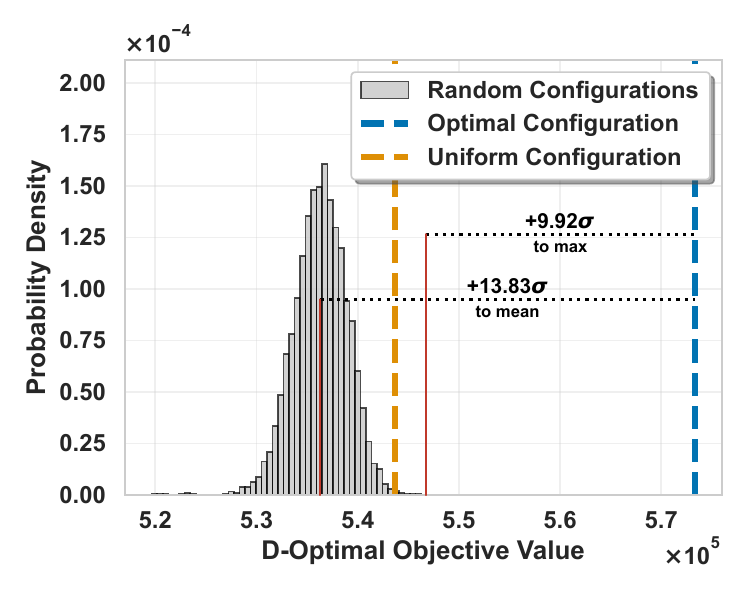}%
    \caption{Histogram of the objective-function values for 10,000 random configurations of 175 sensors each, compared with the greedy-optimal configuration and a uniformly distributed sensor configuration. The greedy-optimal configuration performs significantly better than the random ones, while some random configurations outperform the uniform configuration.}
    \label{fig:oed-hist}
\end{figure}

This hypothetical configuration of 600 sensors, depicted in \Cref{fig:csz-candidate-sensors}, was approximately uniformly distributed over the subduction zone. Given budget and logistical constraints for offshore sensor deployment, it is unlikely that this type of configuration could be deployed in practice \cite{Schmidt2019}. 
More likely, a large-scale sensor network in Cascadia could be of the size of Japan's S-net, which consists of 150 seafloor stations. Thus, using the 600 sensors from~\cite{henneking2025bell} as the candidate set $C$, we performed the optimal sensor selection (\Cref{alg:opt-sens-select-greedy-seq}) for a budget of $B=175$ sensors.

The 600 adjoint PDE solutions of the acoustic--gravity wave equations that are needed to form the shift-invariant p2o map $\vb{F}$ were computed in approximately 500~hours on 512 A100 GPUs, using a finite element implementation of the PDE model in MFEM \cite{anderson2021mfem, tu2026tensor}. After that, forming $\vb{G}^*$ and subsequently $\vb{K}$ took 3~hours on 512 A100 GPUs. Then, the multi-GPU version of \Cref{alg:opt-sens-select-greedy-seq} was run on the candidate matrix $\vb{K}$ using 16 A100 GPUs, taking approximately 12~minutes. All of these computations were performed on Perlmutter.

\begin{figure*}
    \centering
    \includegraphics[width=0.25\textwidth,trim={0 0 65pt 0},clip]
    {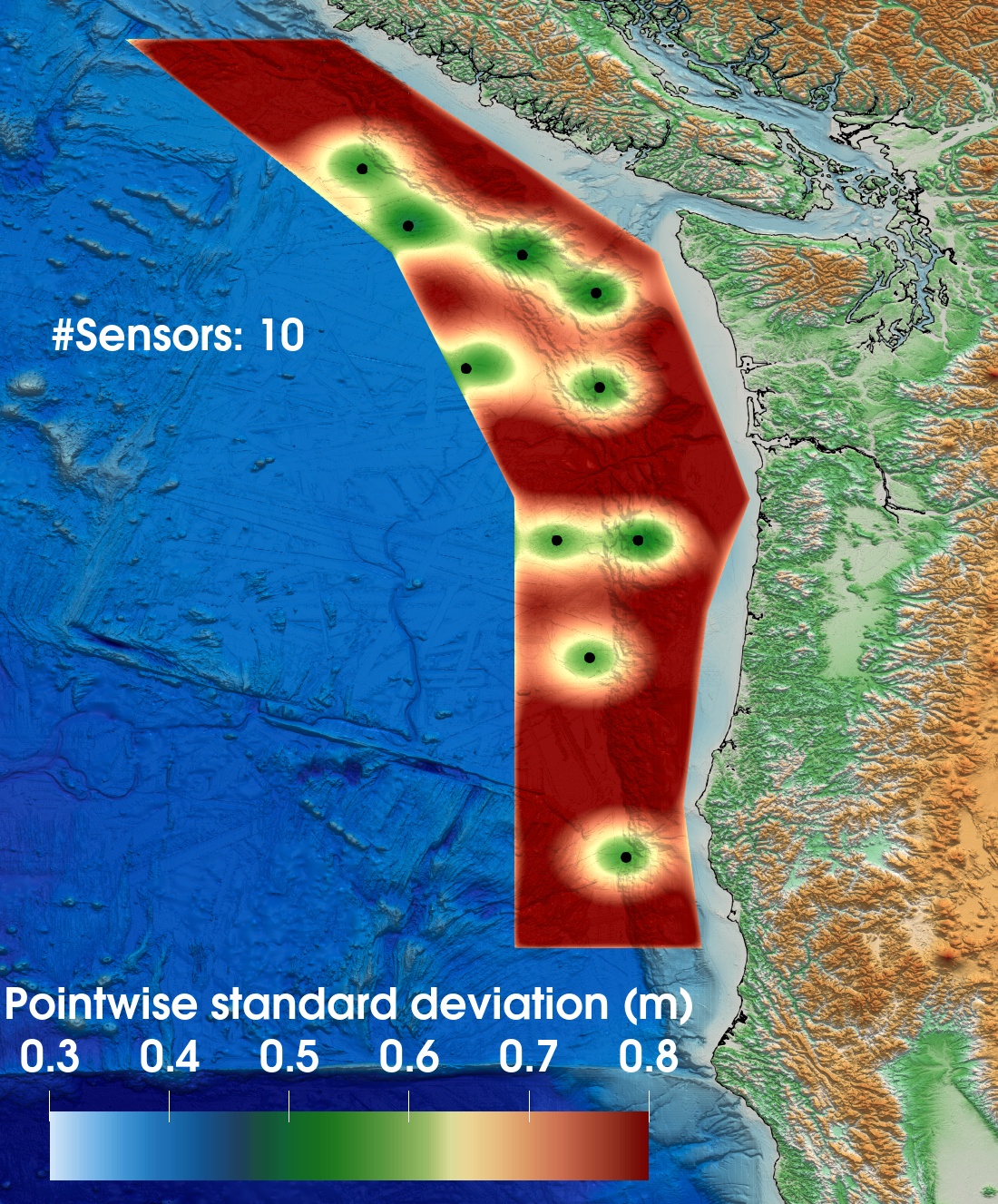}%
    \includegraphics[width=0.25\textwidth,trim={0 0 65pt 0},clip]
    {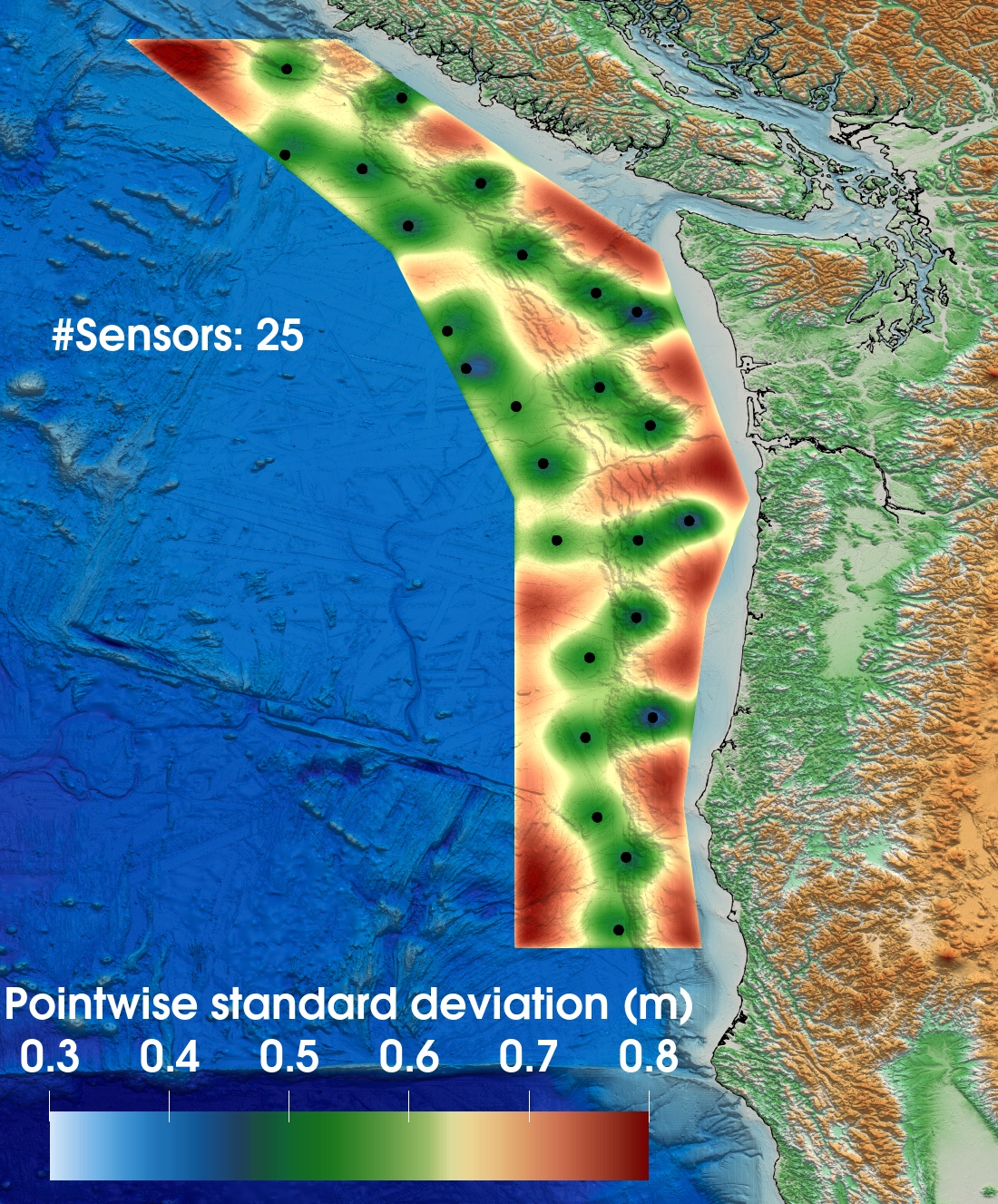}%
    \includegraphics[width=0.25\textwidth,trim={0 0 65pt 0},clip]
    {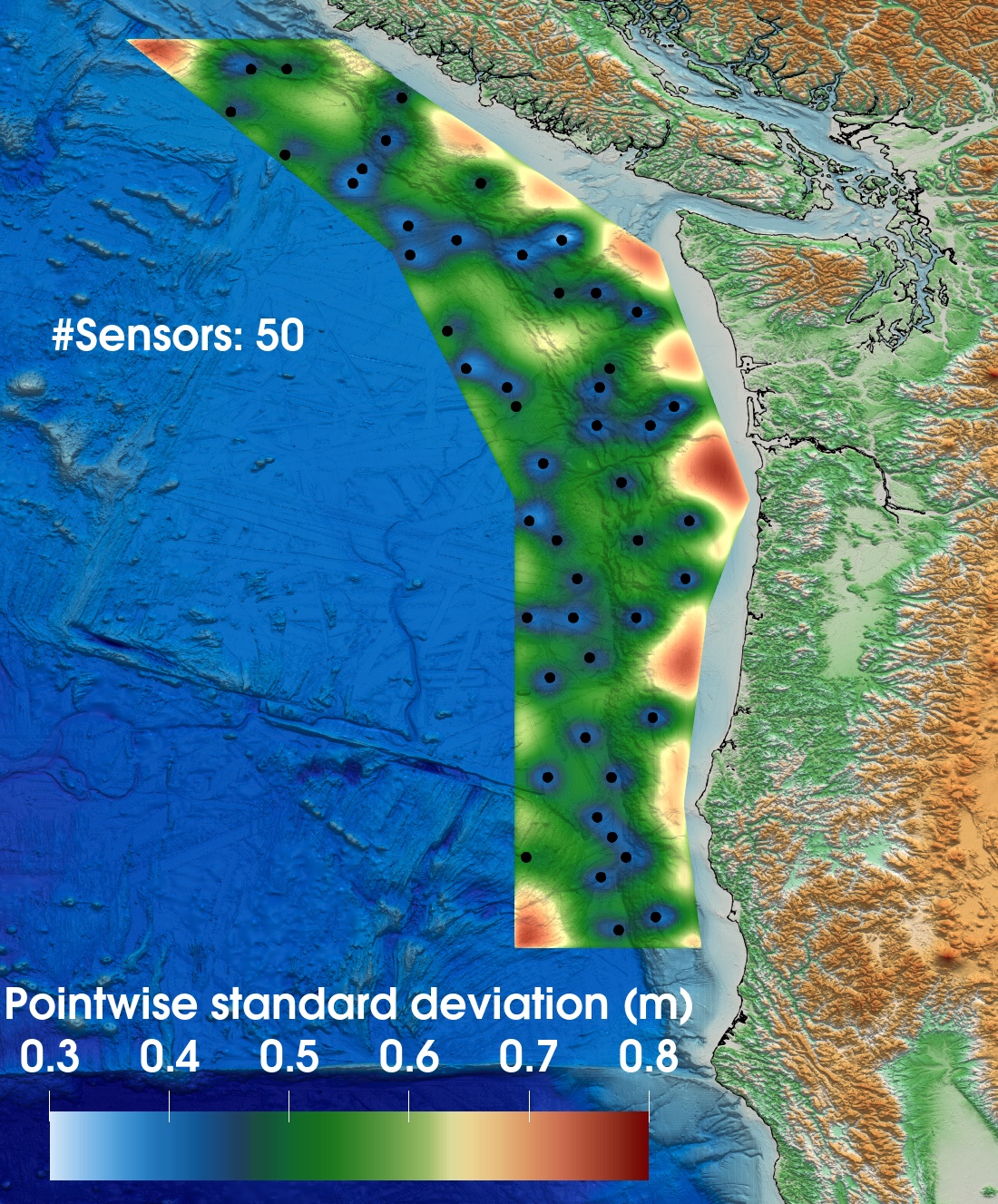}%
    \includegraphics[width=0.25\textwidth,trim={0 0 65pt 0},clip]
    {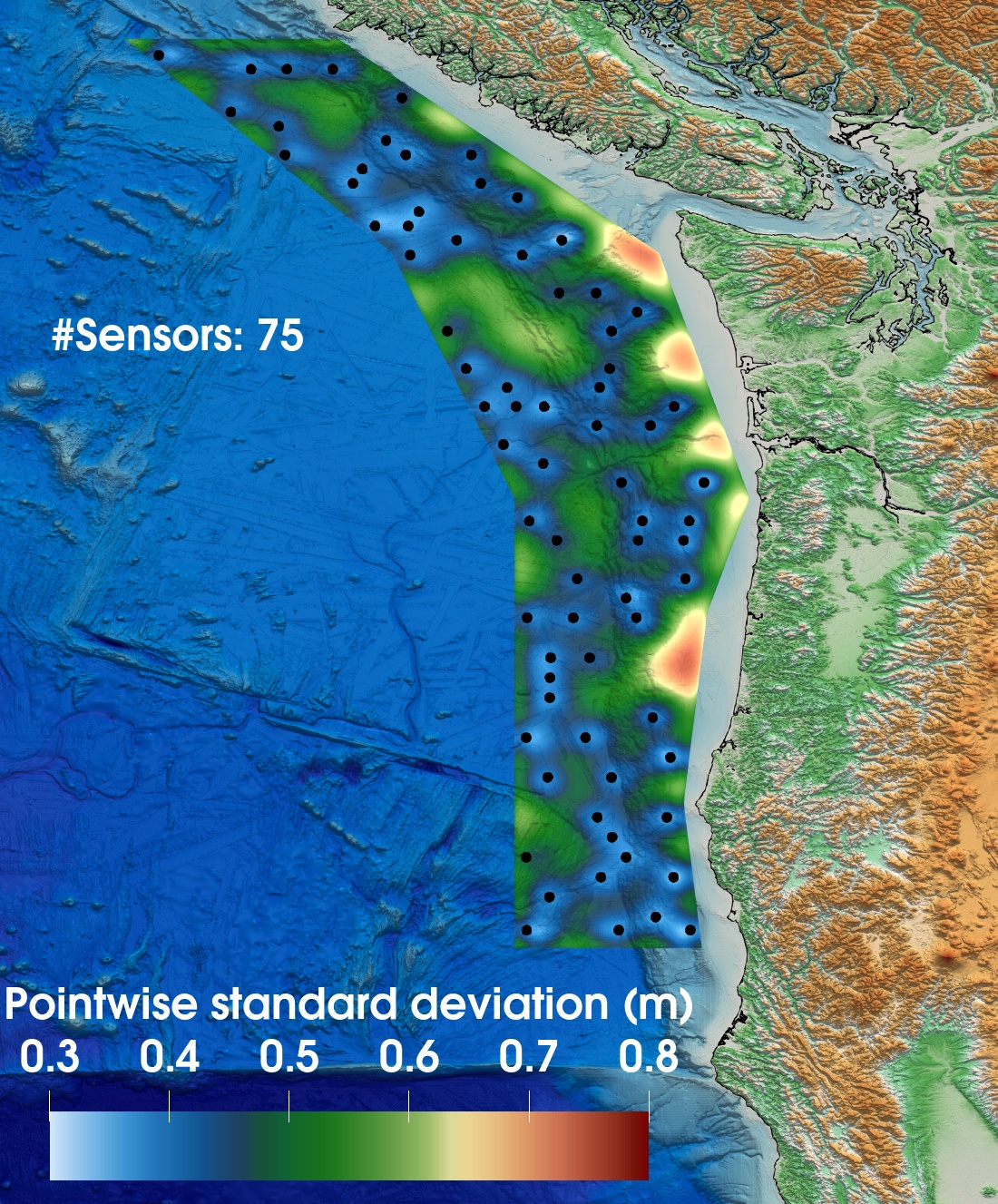}
    \includegraphics[width=0.25\textwidth,trim={0 0 65pt 0},clip]
    {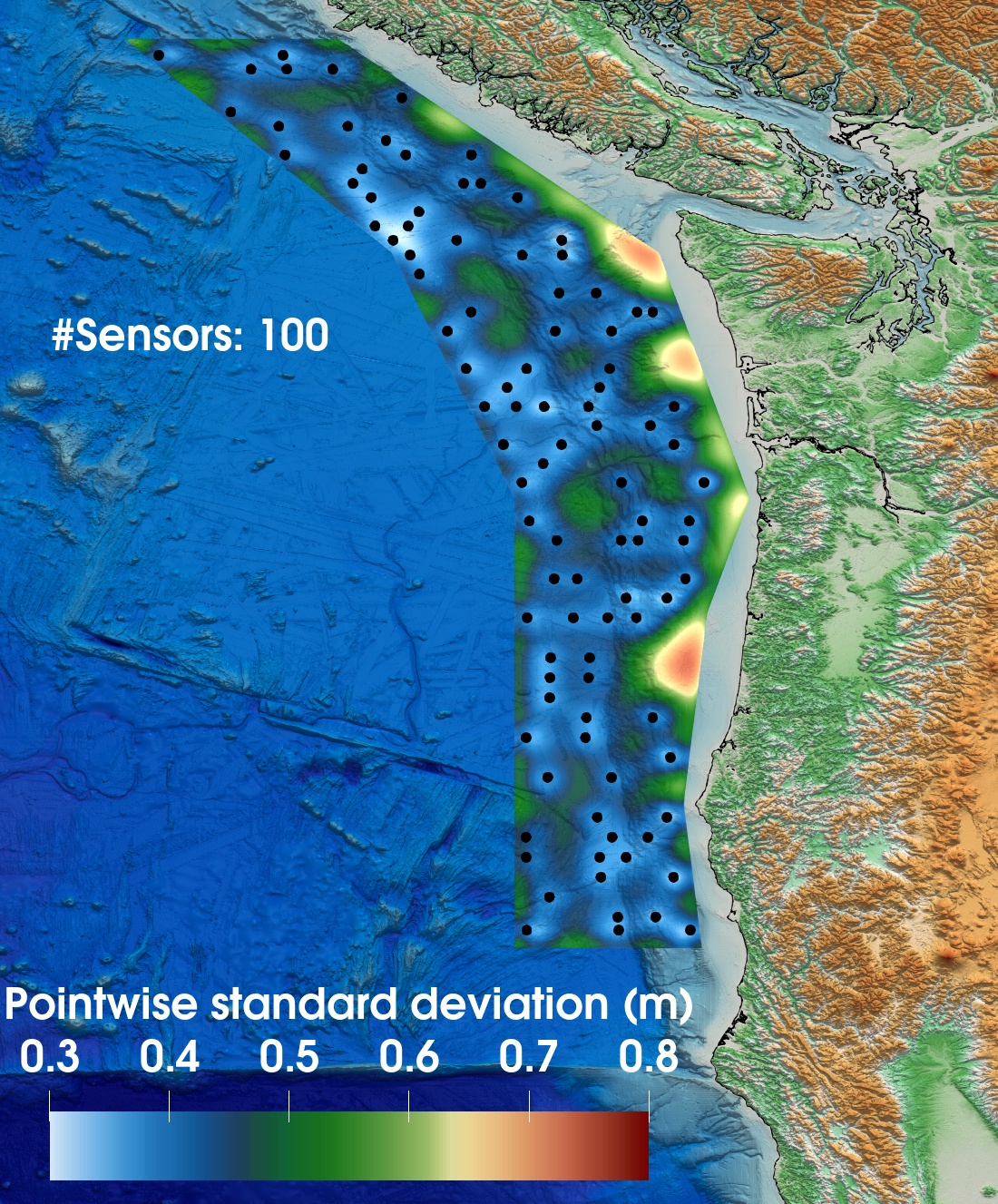}%
    \includegraphics[width=0.25\textwidth,trim={0 0 65pt 0},clip]
    {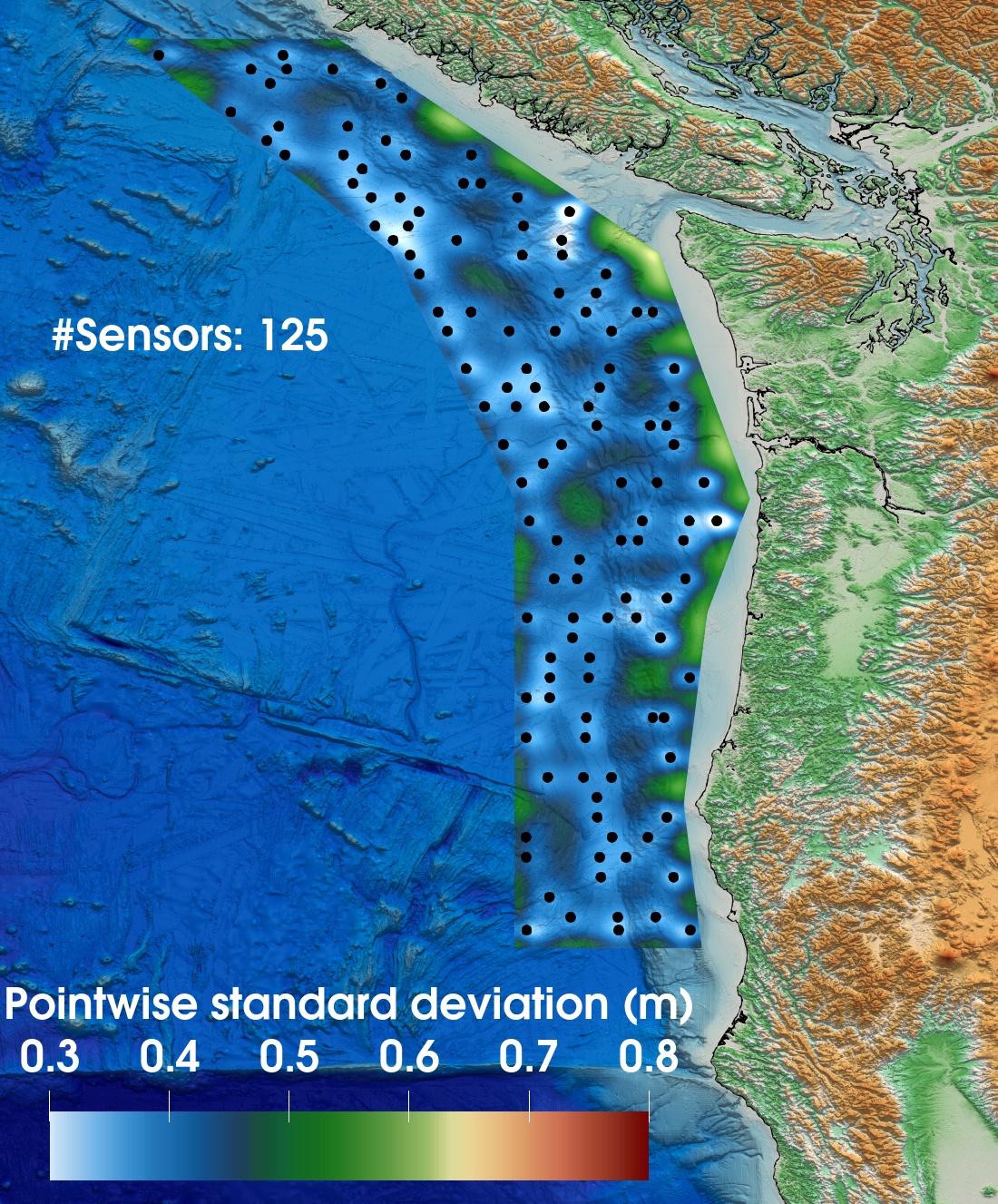}%
    \includegraphics[width=0.25\textwidth,trim={0 0 65pt 0},clip]
    {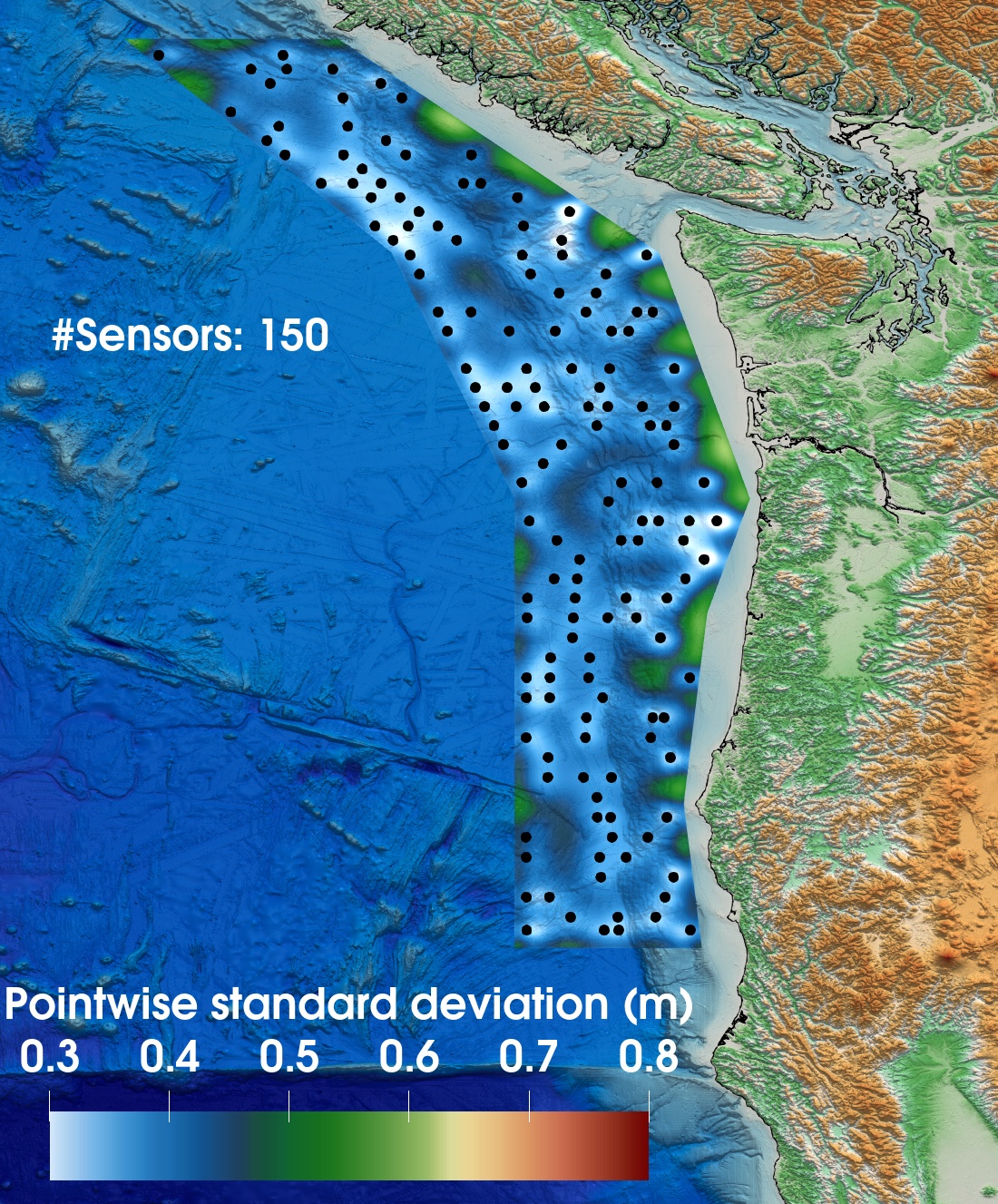}%
    \includegraphics[width=0.25\textwidth,trim={0 0 65pt 0},clip]
    {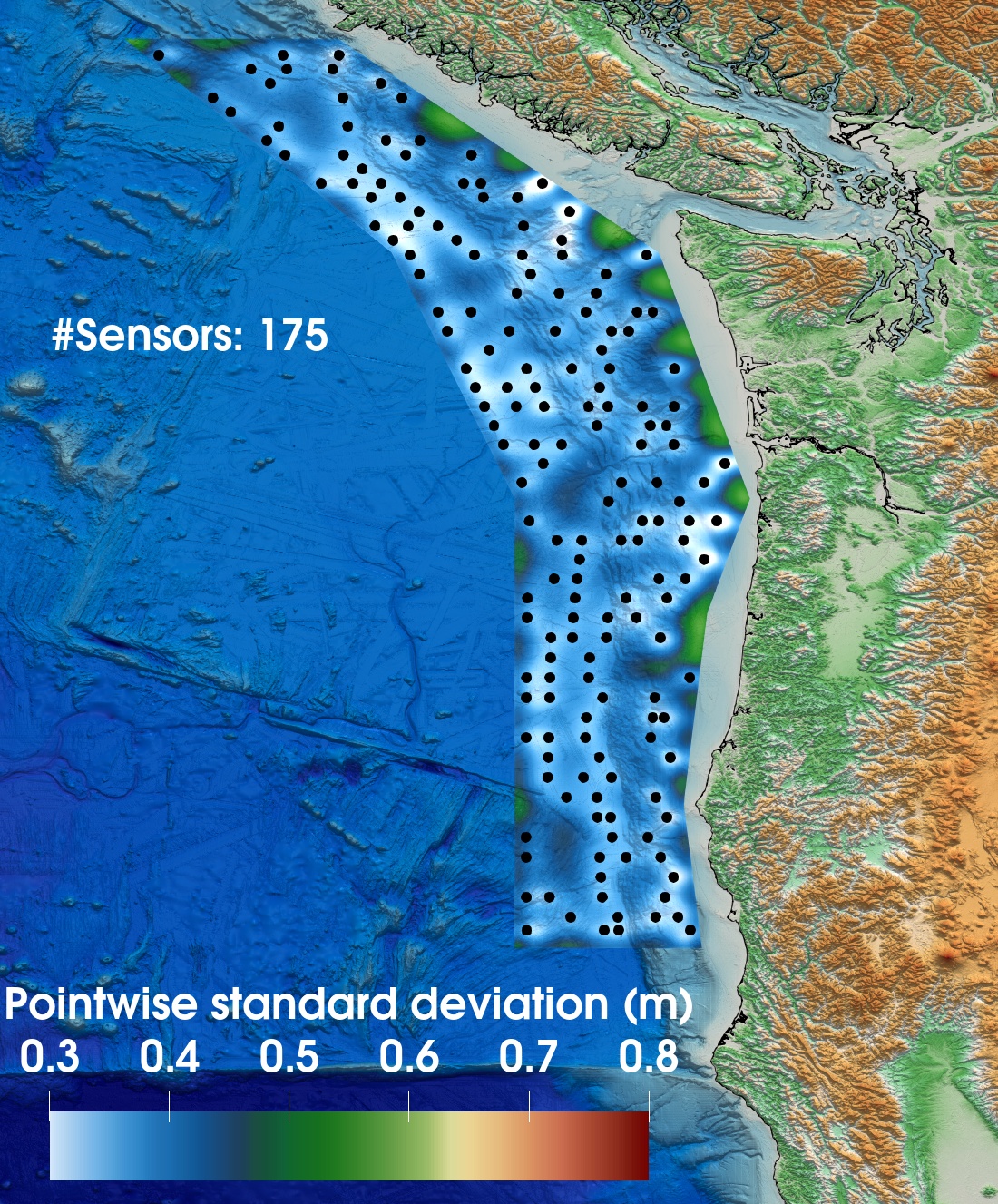}
    \caption{Uncertainties of the inferred seafloor displacement field, illustrated as pointwise standard deviations, for different sensor counts, $|S| \in \{ 10, 25, 50, 75, 100, 125, 150, 175 \}$, during the greedy sensor selection algorithm. 
    Hyperlink to 
    \href{https://www.youtube.com/watch?v=jaXJtnpQWBQ}{\underline{animation}}
    of evolving uncertainty field during greedy sensor selection, at each iteration, $k = 1, 2, \ldots, 175$.}
    \label{fig:oed-selection-uncertainties}
\end{figure*}

As shown in~\Cref{sec:submodularity}, the result of the greedy algorithm will be at least 63\% as good as the global optimizer for the combinatorial optimization problem in~\Cref{eq:d-opt-obj}. Exhaustive evaluation of the global solution space for $B=175$ from 600 candidates is combinatorially intractable, with over $10^{155}$ possible configurations. Instead, the quality of the solution obtained from the greedy algorithm is demonstrated in \Cref{fig:oed-hist}, which compares the values of the objective function $\tilde{\Phi}$ (assuming isotropic noise---see~\Cref{eq:d-opt-obj-computable-id-noise}) for 10,000 random configurations of 175~sensors to that of the greedy-optimal configuration. Additionally, a uniformly-spaced sensor configuration is also evaluated. The greedy-optimal solution has a $z$-score of 13.83 compared to the distribution of random configurations and obtains an objective value 9.92 standard deviations higher than the best-scoring random configuration, while some of the random configurations outperform the uniform configuration.

\Cref{fig:oed-selection-uncertainties} shows the results of \Cref{alg:opt-sens-select-greedy-seq} applied to the Cascadia digital twin with a budget of 175~sensors for the $\vb{K}$ matrix of 600 candidate locations. Specifically, the plots of \Cref{fig:oed-selection-uncertainties} illustrate the reduction in the (time-integrated) posterior uncertainty field, depicted as pointwise standard deviations of the inferred seafloor displacement, as sensors are added iteratively. %
In other words, the plots visualize the information gain from placing sensors according to the greedy optimization algorithm. 
We note that for this linear inverse problem, the posterior covariance $\bfpostcovm$ is independent of the observed data during an earthquake.\footnote{In this linear inverse problem, the uncertainties are informed by the mapping between the seafloor motion parameters and the pressure observables at sensor locations, which is governed by the dynamics of coupled acoustic--gravity wave propagation in the varying-depth ocean. Additional factors that enter the uncertainty calculations are the assumptions on the likelihood, noise model, and prior (see \Cref{sec:background}).} As a result, the sensor selection algorithm seeks to reduce uncertainty in the inferred parameter field over the whole domain uniformly. If we are instead interested in reducing uncertainties in a specific spatiotemporal region (motivated, perhaps, by models of potential rupture scenarios), the masking strategy described in~\Cref{sec:masking} can be used to readily incorporate this information into the sensor selection procedure.

\section{Conclusions}\label{sec:conclusion}

In this paper, we presented a highly scalable, distributed-memory framework for solving extreme-scale Bayesian D-optimal design problems governed by LTI dynamical systems. By recasting the objective from the billion-dimensional parameter space to the $\mc{O}(10^5)$-dimensional data space, we converted the otherwise intractable Bayesian OED problem into a practically computable combinatorial matrix subset selection problem. To make the computation of the Bayesian OED solution efficient at extreme scale, we designed a greedy Schur complement update-based algorithm; this approach eliminates redundant dense matrix factorizations, minimizes memory footprint, and is co-designed to map naturally onto massively parallel GPU architectures. We also developed an MPI-PyTorch implementation of this algorithm that uses a double-buffered pipelined approach to completely overlap I/O with GPU computation. The resulting multi-GPU sensor selection algorithm is performance-portable and exhibits excellent weak and strong scalability over a $256\times$ increase in the number of GPUs on leadership-class supercomputers with varying hardware and filesystem architectures.

We applied this framework to a physics-based, data-driven digital twin for tsunami early warning in the CSZ. The framework selected an optimal 175-sensor network from 600 candidate locations in just 12 minutes on 16 NVIDIA A100 GPUs, solving a Bayesian OED problem that minimizes the uncertainties of the inferred seafloor motion, a parameter field that is discretized with over 1 billion degrees of freedom. To our knowledge, this is the first time that a PDE-constrained Bayesian OED problem has been solved at such a large scale without the use of reduced-order modeling, surrogate modeling, or other approximations of the high-fidelity PDE model. The structure-exploiting design of the algorithm makes the problem tractable; the optimized, pipelined, multi-GPU implementation enables an efficient and scalable solution.

Deploying offshore instrumentation across the Pacific Northwest represents a massive, long-term infrastructure investment. By designing algorithms that exploit problem structure and leverage multi-GPU-accelerated computing, we were able to transform a large-scale Bayesian OED sensor selection problem from a computationally prohibitive task into an agile, iterative design tool. 
This OED framework can be used to rapidly evaluate competing network configurations---targeting different sensor budgets, 
limited sensor deployments in specific areas, 
non-uniform sensor cost models,
or varying instrument noise characteristics---thereby providing the computational tools for guiding future sensor deployments through mathematically rigorous, uncertainty-aware optimization.

\iftoggle{cameraready}{
\section*{Acknowledgments}
This research was supported by DARPA COMPASS grant HR0011-25-3-0242, DOD MURI grant FA9550-24-1-0327, and DOE ASCR grant DE-SC0023171.

This research used resources from the National Energy Research Scientific Computing Center (NERSC) under allocations ALCC-ERCAP0030671, ScienceAtScale DDR-ERCAP0034808 and NESAP DDR-ERCAP0038013. This research used resources of the Oak Ridge Leadership Computing Facility at the Oak Ridge National Laboratory, which is supported by the Office of Science of the U.S.\ Department of Energy under Contract No.\ DE-AC05-00OR22725. 
The authors acknowledge the Texas Advanced Computing Center (TACC) at The University of Texas at Austin for providing computational resources that have contributed to the research results reported within this paper.
}{
}

\bibliographystyle{IEEEtran}
\bibliography{IEEEabrv,paper}

\end{document}